\documentclass[12pt]{article}
\usepackage{amsfonts}
\usepackage{amsmath, amsthm}
\usepackage{epsfig}
\usepackage{algorithm}
\usepackage{algorithmic}
\usepackage{epic}
\usepackage{longtable}

\usepackage[a4paper]{geometry}
\usepackage{enumitem}
\usepackage{dsfont}
\usepackage{rotating}
\usepackage{mathtools}
\usepackage[table,xcdraw]{xcolor}
\usepackage{amsmath,verbatim,color,amssymb,epsfig}
\usepackage{bm}
\usepackage{amsfonts}
\usepackage{epsfig}
\usepackage{changepage}
\usepackage{multirow}
\usepackage{graphicx}
\usepackage{array}
\usepackage[table]{xcolor}
\usepackage{adjustbox,lipsum}
\definecolor{Gray}{gray}{0.80}
\usepackage{lscape}

\usepackage{rotating}
\usepackage{dsfont}
\usepackage[round,semicolon,authoryear]{natbib}
\usepackage[normalem]{ulem}

\usepackage[colorlinks=true,urlcolor=blue,citecolor=purple,linkcolor=blue,bookmarks=true]{hyperref}
\usepackage{caption}

\def\eqx"#1"{{\label{#1}}}
\def\eqn"#1"{{\ref{#1}}}

\makeatletter 
\@addtoreset{equation}{section}
\makeatother  

\def\squarebox#1{\hbox to #1{\hfill\vbox to #1{\vfill}}}
\def\boxit#1{\vbox{\hrule\hbox{\vrule\kern6pt
          \vbox{\kern6pt#1\kern6pt}\kern6pt\vrule}\hrule}}

\newtheorem{thm}{Theorem}
\newtheorem{lem}{Lemma}

\newcolumntype{L}[1]{>{\raggedright\let\newline\\\arraybackslash\hspace{0pt}}m{#1}}
\newcolumntype{C}[1]{>{\centering\let\newline\\\arraybackslash\hspace{0pt}}m{#1}}
\newcolumntype{R}[1]{>{\raggedleft\let\newline\\\arraybackslash\hspace{0pt}}m{#1}}

\usepackage{booktabs,multirow}
\usepackage{tikz}
\usetikzlibrary{shapes}

\usepackage{threeparttable}

\newcommand{\beginsupplement}{%
        \renewcommand{\thetable}{S\arabic{table}}%
        \renewcommand{\thefigure}{S\arabic{figure}}%
    \renewcommand{\thesection}{S\arabic{section}}%
    \renewcommand{\thesubsection}{S\arabic{section}.\arabic{subsection}}%
    }

\begin{document}

\baselineskip=24pt
\begin{center}
{\Large \bf Randomized Optimal Selection Design for Dose Optimization}
\end{center}
\begin{center}
{\bf Shuqi Wang, Ying Yuan$^*$, Suyu Liu$^{**}$}
\end{center}

\begin{center}
Department of Biostatistics, The University of Texas MD Anderson Cancer Center\\
Houston, TX, USA.\\
{*Email: yyuan@mdanderson.org;  $\qquad$ **Email: syliu@mdanderson.org}\\
\end{center}

\begin{center}
\vskip 0.2 true cm

{\textbf{Abstract}}
\end{center}
\vskip 0.15 true cm
The U.S. Food and Drug Administration (FDA) launched Project Optimus to shift the objective of dose selection from the maximum tolerated dose to the optimal biological dose (OBD), optimizing the benefit-risk tradeoff. One approach recommended by the FDA's guidance is to conduct randomized trials comparing multiple doses. In this paper, using the selection design framework \citep{Simon1985}, we propose a \underline{r}andomized \underline{o}ptimal \underline{se}lection (ROSE) design, which minimizes sample size while ensuring the probability of correct selection of the OBD at prespecified accuracy levels. The ROSE design is simple to implement, involving a straightforward comparison of the difference in response rates between two dose arms against a predetermined decision boundary. We further consider a two-stage ROSE design that allows for early selection of the OBD at the interim when there is sufficient evidence, further reducing the sample size. Simulation studies demonstrate that the ROSE design exhibits desirable operating characteristics in correctly identifying the OBD. A sample size of 15 to 40 patients per dosage arm typically results in a percentage of correct selection of the optimal dose ranging from 60\% to 70\%. A user-friendly software for implementing ROSE designs is available at www.trialdesign.org.

\vskip 0.15 true cm
\noindent{KEY WORDS:} Dose optimization; Optimal design; Randomization; Selection design.

\baselineskip=24pt

\newpage

\section{Introduction} \label{sec:Introduction}

The conventional dose-finding paradigm, originally motivated by cytotoxic chemotherapies based on maximum tolerated dose (MTD), is often unsuitable for targeted molecules or immunotherapies \citep{Ratain2014, Yan2018}. These novel therapies exhibit different characteristics from cytotoxic agents; for example, their efficacy often plateaus before reaching the MTD, or the MTD itself may be unattainable \citep{Shah2021}. This suggests that patients may be given a lower dose with similar efficacy and potentially fewer side effects (\citealp{Jain2010, Sachs2016, Zirkelbach2022}). Consequently, the U.S. Food and Drug Administration (FDA) launched Project Optimus \citep{FDA2022} and released draft guidance \citep{FDA2023-dose-optimization} to shift the dose-finding paradigm from identifying the MTD to determining the optimal biological dose (OBD) that maximizes a toxicity-response tradeoff or provides the desired therapeutical effect while minimizing toxicity.

In the FDA's guidance for dose optimization, a recommended approach is to conduct a randomized, parallel dose-response trial to compare multiple doses after completing conventional phase I dose escalation and identifying the MTD. \citet{Hoering2011} proposed a two-step dose-finding trial in which step 1 employs a traditional phase I trial design to determine the MTD, and step 2 utilizes a modified phase II selection design for multiple doses to determine the OBD based on both efficacy and toxicity. 
\citet{UBOIN2019} developed a utility-based Bayesian optimal interval (U-BOIN) design, which first performs dose escalation in stage 1 and then randomized patients among admissible doses in stage 2 to identify the OBD based on utility. \citet{DROID2023} proposed a dose-ranging approach to optimizing dose (DROID), where patients are randomized to multiple dose for comparison and identification of the OBD by modeling dose-response curves.  \citet{DAngelo2024} proposed U-MET design to identify the OBD by comparing utility between randomized doses. None of these methods, however, provide a concrete way to determine the sample size and ensure that the probability of correct selection (PCS) of the OBD reaches certain prespecified accuracy levels. \citet{MERIT2024} proposed a multiple-dose randomized trial (MERIT) design for dose optimization that provides a rigorous algorithm for sample size determination based on the hypothesis testing framework. However, that method relies on modified definitions of type I error and power, which may be unfamiliar to practitioners, and focuses on identifying the OBD-admissible doses rather than directly addressing the accuracy of selecting the OBD.  The objective of this paper is to fill these gaps by proposing a design that provides a rigorous way to determine the sample size while explicitly ensuring that the PCS reaches prespecified accuracy levels.

A key challenge in designing randomized dose optimization trials is the often unrealistically large sample sizes required when using the conventional method based on testing the null hypothesis of therapeutic equivalence. This challenge is noted in the FDA guidance, which states: ``The (randomized dose optimization) trial should be sized to allow for sufficient assessment of safety and antitumor activity for each dosage. The trial does not need to be powered to demonstrate statistical superiority of a dosage or statistical non-inferiority among the dosages using Type I error rates which would be used in registrational trials."

The selection design proposed by \citet{Simon1985} offers a potential solution to this challenge. Based on statistical selection theory and relying on ranking, the selection design requires a substantially smaller sample size than the conventional method based on hypothesis testing. For example, when comparing two treatments, the selection design requires 21 patients per treatment arm to achieve a 90\% probability of correctly selecting the superior treatment when the efficacy of the two treatments is 10\% and 25\%. In contrast, standard power calculations would require 143 patients per arm to detect the difference between the two treatments with 90\% power at a two-sided 5\% significance level, based on Fisher's exact test.

Unfortunately, the selection design proposed by \citet{Simon1985} cannot be directly applied to dose optimization for two main reasons. First, the selection design aims to identify the most efficacious treatment among several independent treatments, focusing on the pick-the-winner rule—selecting the dose with the highest efficacy rate. In contrast, the goal of dose optimization is to identify the dose that achieves the optimal risk-benefit balance. Even if a higher dose has a higher efficacy rate, a lower dose may be preferable if its efficacy rate is sufficiently close to that of the higher dose, due to its better safety profile.

Additionally, the selection design assumes that the treatments under comparison are independent and their response rates are unordered. In contrast, randomized dose optimization trials consider ordered doses of the same drug, where the toxicity and efficacy of the higher dose should not be lower than those of the lower dose. This order constraint requires special statistical considerations, as discussed later.
 
To address these challenges, we propose a Randomized Optimal SElection (ROSE) design. We focus on the two-dose randomized dose optimization trial, as most practical dose optimization trials involve two doses, typically the MTD and one dose lower than the MTD. Under the ROSE design, the high dose is selected as the OBD only if the response improvement of the high dose compared to the low dose reaches a certain threshold; otherwise, the lower dose is selected.  The threshold is optimized to minimize the sample size while ensuring a desirable level of accuracy in correctly selecting the OBD. The ROSE design is easy to implement and flexible, allowing for different accuracy requirements in correctly selecting the high dose and the low dose.

The remainder of the paper is organized as follows. In Section \ref{sec:Method}, we introduce optimal selection design, including the dose selection rule and algorithms for determining the decision boundary and sample size. Extensive simulations are conducted to evaluate the operating characteristics of the proposed design in Section \ref{sec:simulation}. We provide a discussion in Section \ref{sec:Discussion}.

\section{Method} \label{sec:Method}
\subsection{Dose Selection Rule} \label{subsec:dose_selection_rule}
Consider a two-dose randomized dose optimization trial, where a total of $2 \times n$ patients are equally randomized to two candidate doses $d_L$ and $d_H$. 
The scenario with an unequal randomization ratio will be discussed later. The doses $d_L$ and $d_H$ may be determined based on clinical outcomes, pharmacokinetic/pharmacodynamic (PK/PD) data, or other biological variables from previous dose escalation trials. Typically, the high dose $d_H$ is the MTD identified in a previous study, while $d_L$ is a pre-specified lower dose. Let $Y_{i,j}$ denote the binary response endpoint for the $i$-th patient treated with dose $d_j$, where $j =L, H$, $i=1,\dots,n$. Examples include tumor objective response or biological response, indicating whether a biomarker (e.g., receptor occupancy, ctDNA clearance) reaches a specified minimum activity level.  Let $p_j$ denote the true response rate for dose $d_j$. We assume $Y_{i,j}$ are independent and identically distributed (i.i.d.), and follow a Bernoulli distribution with $\Pr(Y_{i,j}=1)=p_j$.  As is often the case in practice, the dose-toxicity curve is assumed to be monotonically increasing with the dose, while the dose-response curve is assumed to be monotonically non-decreasing.  This implies that the toxicity of $d_L$ is always lower than that of $d_H$, whereas the efficacy of $d_L$ can be lower or equal to that of $d_H$ (i.e., $p_L \leq p_H$). 

The aim of dose optimization is to select the optimal dose that maximizes the benefit-risk tradeoff, rather than the dose with the highest response rate. The benefit-risk tradeoff can be modeled explicitly, for example, using a utility, or implicitly. In the utility approach, each possible toxicity-efficacy outcome is assigned a utility score to reflect its clinical desirability \citep{BOIN12}. This approach is highly flexible and statistically rigorous, but it is more involved and requires the specification of a utility score for each possible outcome. Here, we focus on an implicit approach based on the following OBD selection rule: 
\begin{itemize}
\item If $\widehat{p}_H - \widehat{p}_L \leq \lambda$, we select $d_L$ as the OBD;
\item Otherwise, i.e., $\widehat{p}_H - \widehat{p}_L > \lambda$, we select $d_H$ as the OBD.
\end{itemize}
where $\widehat{p}_j = \frac{1}{n}\sum_{i=1}^n Y_{i,j}$, $j=L, H$, is the sample estimate of $p_j$, and $\lambda$ is a decision boundary. Although this rule seems to depend only on efficacy (i.e., $\widehat{p}_j$), it implicitly considers the toxicity-efficacy tradeoff. The decision rule accounts for the fact that $d_L$ is safer. This is reflected in the condition that if the response rate of $d_H$ is not higher than that of $d_L$ by a certain margin (i.e., $\widehat{p}_H - \widehat{p}_L \leq \lambda$), then $d_L$ is preferred and selected.

The value of $\lambda$ determines the operating characteristics of the design. As $\lambda$ increases, the probability of selecting $d_L$ increases, while the probability of selecting $d_H$ decreases. In what follows, we will describe a procedure to identify the optimal $\lambda$ that minimizes the sample size while ensuring the PCS of the  OBD at prespecified levels.

\subsection{Optimal Decision Boundary and Sample Size} \label{subsec:boundary_sample size}
As the primary objective of randomized dose optimization trials is to identify the OBD, it is critical that the trial design ensures a desirable level of PCS of the OBD. To proceed, we consider two representative scenarios of interest:
\begin{eqnarray} \nonumber \label{eqn:hypothesis}
    \begin{aligned}
     & \mathcal{S}_L: p_H - p_L=0 \\
     & \mathcal{S}_H: p_H - p_L \geq \delta.\\
    \end{aligned}
\end{eqnarray}
$\mathcal{S}_L$ represents that the dose-efficacy curve has plateaued after $d_L$, thus $d_L$ is the OBD because it is safer. $\mathcal{S}_H$ represents the case where the dose-efficacy curve is still in the increasing range from $d_L$ to $d_H$. More precisely, $p_H$ is better than $p_L$ by a significant margin $\delta$, thus $d_H$ should be selected as the OBD. The rationale for selecting $d_H$ is that since $d_L$ and $d_H$ have been previously evaluated for safety in phase I trials, they are presumed to not exceed the MTD and are highly likely to be acceptable in terms of dose-limiting toxicity or severe adverse events. Therefore, when $d_H$ is substantially more efficacious than $d_L$, we should select $d_H$ as the OBD, considering that the lack of efficacy is the primary factor contributing to the failure of drug development. This also provides the interpretation of $\delta$, which represents the treatment gain from $d_H$ that outweighs the potential safety benefit of $d_L$. 

The value of $\delta$ should be chosen based on clinical and statistical considerations. In general, a smaller $\delta$ will require a larger sample size, given the same targeted PCS. As the sample size for randomized dose optimization trials is typically small, we should refrain from setting $\delta$ at a very small value.  One practical approach is to elicit the initial value of $\delta$ from clinicians by asking them under what treatment difference they would choose $d_H$. Based on that initial value, we generate the operating characteristics of the design (e.g., sample size and PCS) over a grid of $\delta$ around that initial value, using the software described later. The most appropriate value of $\delta$ can be determined jointly by clinicians and statisticians, based on reviewing the operating characteristics, to balance the competing aspects of the trials (e.g., PCS and sample size).  This can be an iterative calibration process. 

Based on the simulation studies presented later, $\delta=0.1$ or 0.15 is generally  a reasonable choice. While these values may appear large, the actual decision boundary $\lambda$ used to choose $d_L$ or $d_H$ is substantially smaller — typically around half of  $\delta$, as described later. For example, when  $\delta=0.1$, $\lambda$ is often approximately 0.05,  meaning that  $d_H$ will be selected when its observed response rate exceeds that of  $d_L$ by 0.05 or more.  

It is important not to confuse $\delta$ with $\lambda$. For example, it is incorrect to interpret ${\cal S}_H$ as meaning that $d_H$ is selected only when the response rate difference between the two doses exceeds $\delta$. The margin $\delta$ represents a “strong” case in which $d_H$ is clearly preferred, and thus a certain PCS should be assured. In contrast, $\lambda$ is the decision boundary chosen to achieve that assured PCS. The purpose of specifying $\mathcal{S}_L$ and $\mathcal{S}_H$ is not to conduct hypothesis testing, but rather to define cases of particular interest—namely, $\mathcal{S}_L$ and $\mathcal{S}_H$ represent the scenarios in which $d_L$ and $d_H$ are the OBD, respectively—and to ensure that the design performs well under these clear-cut anchoring cases.

We now discuss how to determine the optimal decision boundary $\lambda$ to ensure PCS at prespecified levels. Given $\mathcal{S}_L$ and $\mathcal{S}_H$, the PCS under these scenarios of interest are respectively given by
\begin{align}
\beta_L(\lambda) &= \Pr(\text{Select } d_L \mid \mathcal{S}_L) \nonumber \\ \nonumber
                 &= \Pr(\widehat{p}_H - \widehat{p}_L \leq \lambda \mid \mathcal{S}_L) \nonumber \\
                 &\approx \Phi\left( \frac{\sqrt{n}\lambda}{\sqrt{2 p_H(1-p_H)}} \mid \mathcal{S}_L \right),  \nonumber \\
\intertext{and} 
\beta_H(\lambda) &= \Pr(\text{Select } d_H \mid \mathcal{S}_H) \nonumber \\ \nonumber
                 &= \Pr(\widehat{p}_H - \widehat{p}_L > \lambda \mid \mathcal{S}_H) \nonumber \\
                 &\approx 1 - \Phi\left( \frac{\sqrt{n}(\lambda - \delta)}{\sqrt{(p_H - \delta)(1 - p_H +\delta) + p_H (1 - p_H)}} \mid \mathcal{S}_H \right). \nonumber
\end{align}
where $\Phi(\cdot)$ is the cumulative distribution function (CDF) of a $N(0,1)$ distribution. $\beta_H(\lambda)$ is derived under $\mathcal{S}_H$ with $p_L = p_H - \delta$. Given $p_H$ and $\delta$, ensuring adequate PCS under this setting guarantees sufficient PCS for all $p_L < p_H - \delta$. The detailed derivation is provided in Supplementary Material \ref{sec:approx_PCS}.

The ROSE design controls the PCS under $\mathcal{S}_L$ and $\mathcal{S}_H$ at prespecified levels of $\alpha_L$ and $\alpha_H$, respectively, by choosing an appropriate decision boundary $\lambda$, i.e., 
\begin{equation} \nonumber
\beta_L(\lambda) \geq \alpha_L ~\,\, \mathrm{and}\,\,~ \beta_H(\lambda) \geq \alpha_H.
\end{equation}
Since the selection is made between two candidate doses, both $\alpha_L$ and $\alpha_H$ should greater than $0.5$. As demonstrated in the numerical study presented later, setting $\alpha_L$ and $\alpha_H$ within the range $[0.6, 0.7]$ is a reasonable choice, consistent with the accuracy levels achieved in typical MTD dose-finding trials and resulting in  a feasible sample size for early-phase studies. These parameters can be adjusted according to the desired accuracy and sample size considerations specific to the trial.

Given a fixed $n$, this results in the following solution:
\begin{equation} \label{intervalsoluation}
\lambda \in \left(\lambda_L \equiv \frac{\sqrt{2 p_H(1-p_H)}}{\sqrt{n}}\Phi^{-1}(\alpha_L), \, \lambda_H \equiv \frac{\sqrt{(p_H-\delta)(1-p_H+\delta) + p_H(1-p_H)}}{\sqrt{n}}\Phi^{-1}(1-\alpha_H) + \delta \right)
\end{equation}
when $\lambda_L\le\lambda_H$. When $\lambda_L > \lambda_H$,  $\lambda$ has no solution, which means that the PCS under $\mathcal{S}_L$ and $\mathcal{S}_H$ cannot simultaneously reach their respective desired levels.

Next, we discuss the optimization of the design by determining the minimal value of $n$ such that the solution for $\lambda$ exists. In other words, we aim to identify the design that satisfies the PCS requirement with the minimal sample size. The optimization is provided by the following results, and the proof of theorem \ref{theorem1-samplesize} is provided in Supplementary Material \ref{Sup: proof-theorem1-samplesize}. 

\begin{thm} \label{theorem1-samplesize}
Sample size \(n\) is minimized when \(\lambda_L = {\lambda_H}\), which is given by     \begin{equation} \label{eqn:sample_size}
        n = \left(\frac{\sqrt{2 p_H(1-p_H)}\Phi^{-1}(\alpha_L) - \sqrt{(p_H-\delta)(1-p_H+\delta) + p_H(1-p_H)}\Phi^{-1}(1-\alpha_H) }{\delta}\right)^2.
    \end{equation}
\end{thm}

\begin{lem}
The decision boundary $\lambda$ corresponding to the minimal $n$ provided in Theorem  \ref{theorem1-samplesize}  is given by
\begin{equation} \label{eqn:decision_boundary}
\lambda = \frac{ \delta \sqrt{2 p_H(1-p_H)}\Phi^{-1}(\alpha_L)}{\sqrt{2 p_H(1-p_H)}\Phi^{-1}(\alpha_L) - \sqrt{ (p_H-\delta)(1-p_H+\delta) + p_H(1-p_H)}\Phi^{-1}(1-\alpha_H)}.
\end{equation}
\end{lem}

The sample size provided by Theorem \ref{theorem1-samplesize} may not be an integer. In practice, $n$ can be rounded to its ceiling, $\lceil n \rceil$. The question then arises: do we need to recalculate $\lambda$ based on $\lceil n \rceil$? The answer is no. As shown in the proof of Theorem \ref{theorem1-samplesize}, $\lambda_L$ is a decreasing function of $n$, whereas $\lambda_H$ is an increasing function of $n$, due to the fact that $\Phi^{-1}(\alpha_L)>0$ and $\Phi^{-1}(1-\alpha_H)< 0$, given that $\alpha_L >0.5$ and $\alpha_H > 0.5$. Since $\lceil n \rceil \geq n$, the interval $(\lambda_L, \lambda_H)$ computed using the rounded sample size $\lceil n \rceil$ will encompass the exact decision boundary $\lambda$ derived from equation (\ref{eqn:decision_boundary}) based on the unrounded sample size $n$. Of note, as $\lceil n \rceil \geq n$, the resulting $\lambda_L \neq \lambda_H$ and the solution will be an interval as provided by equation (\ref{intervalsoluation}). Since any value within this interval satisfies the PCS requirement, we can simply use $\lambda$ provided by equation (\ref{eqn:decision_boundary}) for dose selection.  

Tables \ref{tab:Bounderies_sample_size_delta_0.1} and \ref{tab:Bounderies_sample_size_delta_0.15} present the decision boundaries and minimum sample sizes per dose arm for various parameter configurations, with $\delta = 0.1$ and 0.15, respectively. Corresponding results for $\delta = 0.05$ are provided in Supplementary Material \ref{sec:ROSE_delta_0.05}. 
The left side of each table represents the one-stage ROSE design without interim monitoring. As expected, reaching a high PCS requires a larger sample size. For example, when $p_H=0.3$ and $\delta=0.1$, a sample size of 11 patients per dose arm is required to achieve  $\alpha_L = \alpha_H = 0.6$, whereas 44 patients per dose arm are required to achieve $\alpha_L = \alpha_H = 0.7$.  Moreover, increasing the value of $\delta$ reduces the sample size needed to attain the same PCS. For instance, under $p_H = 0.3$, achieving $\alpha_L = \alpha_H = 0.7$ requires 44 patients per arm when $\delta = 0.1$, but only 19 patients per arm when $\delta = 0.15$.  Furthermore, given the same desired accuracy levels for high and low doses, i.e., $\alpha_L = \alpha_H$, the decision boundary $\lambda$ is close to $\delta/2$.

The results above assume equal randomization between $d_H$ and $d_L$. A more flexible version of ROSE that accommodates an allocation ratio of $C:1$ between $d_H$ and $d_L$ is presented in Supplementary Material \ref{sec:allocarion_ratio_C}.
\subsection{Two-stage ROSE Design with Interim Monitoring} \label{subsec:two_stage_ROSE}
Interim monitoring can be incorporated into the ROSE design to allow early stopping of the trial if the interim data provide sufficient evidence to support dose selection. This avoids exposing more patients to suboptimal dose, reduces the sample size, and accelerates the development timeline. As the sample size for randomized dose optimization trials is typically small to moderate and such trials often aim to collect as much data as possible to facilitate the design and decision of subsequent pivotal trials, one interim monitoring is sufficient for most applications. Therefore, we focus on a two-stage ROSE design with an interim analysis. 

Let $\omega$ denote the sample size fraction at which the interim analysis is planned. For example, when $\omega=0.5$, the interim analysis is conducted when half of the patients, i.e. $n/2$,  have been enrolled and their efficacy has been evaluated for each dose. Let $n_1$ denote the number of patients enrolled in each dose in the interim monitoring point, i.e., $n_1 = \omega n$. The estimated response rate for dose $j$ at the interim analysis is denoted by $\widehat{p}_{1,j}$, where $\widehat{p}_{1,j} =\frac{1}{n_1} \sum_{i=1}^{n_1}Y_{i,j}$, $j={L,H}$. Recall that $\widehat{p}_j$ represents the estimated response rate for dose $j$ based on total sample size $n$. The two-stage ROSE design is given as follows:
\begin{enumerate}[label=Step \arabic*., leftmargin=*]
    \item[Step 1.] Enroll $n_1$ patients for each dose arm:
    \begin{itemize} [leftmargin=*]
        \item If $\widehat{p}_{1,H} - \widehat{p}_{1,L} >  \lambda_1$, we terminate the trial early and select $d_H$ as the OBD;
        \item  Otherwise, i.e., $\widehat{p}_{1,H} - \widehat{p}_{1,L} \leq \lambda_1$, we continue the trial and proceed to step 2.
    \end{itemize}
    \item[Step 2.] Enroll additional $(n - n_1)$ patients for each dose arm:
    \begin{itemize}[leftmargin=*]
        \item If $\widehat{p}_H - \widehat{p}_L \leq \lambda$, we select $d_L$ as the OBD;
        \item Otherwise, i.e., $\widehat{p}_H - \widehat{p}_L > \lambda$, we select $d_H$ as the OBD.
    \end{itemize}
\end{enumerate} 
Here, $\lambda_1$ and $\lambda$ are decision boundaries for interim analysis and final analysis, respectively, with $0 \le \lambda \le \lambda_1 \le 1$. This implies that the trial will only be terminated early if there is strong evidence indicating that the efficacy of $d_H$ is superior to that of $d_L$, in which case it would be ethically concerning to continue treating patients at $d_L$. 

We do not consider early selection of $d_L$ when $\widehat{p}_{1,H}$ and $\widehat{p}_{1,L}$ are close, because the small interim sample size may lead to unreliable conclusions. In addition, although certain endpoints (e.g., response) are chosen to define the decision rule for design purposes, dose optimization trials are intended to collect a broader range of data, such as pharmacokinetics/pharmacodynamics (PK/PD), tolerability, and quality of life. The final dose selection will therefore depend on both the design recommendation and the totality of evidence to assess the risk–benefit tradeoff. Therefore, when two arms show similar efficacy, there is no ethical justification to stop enrollment in either arm; instead, it is valuable to continue the trial and collect additional data.

Define the PCS under two scenarios of interest as $\beta_L(\lambda_1, \lambda)$ and $\beta_H(\lambda_1, \lambda)$, respectively, expressed as:
\begin{eqnarray} \label{eqn:power_interim_H0} \nonumber
    \begin{aligned}
    \beta_L(\lambda_1, \lambda) = \Pr(\widehat{p}_{1,H} - \widehat{p}_{1,L} \leq \lambda_1, \,\,  \widehat{p}_{H} - \widehat{p}_{L} \leq \lambda|\mathcal{S}_L) \\
    \end{aligned}
\end{eqnarray}
and
\begin{eqnarray} \label{eqn:power_interim_H1} \nonumber
    \begin{aligned}
    \beta_H(\lambda_1, \lambda) =  \Pr(\widehat{p}_{1,H} - \widehat{p}_{1,L} > \lambda_1 \mid \mathcal{S}_H) + \Pr(\widehat{p}_{1,H} - \widehat{p}_{1,L} \leq \lambda_1, \,\,\widehat{p}_{H} - \widehat{p}_{L} > \lambda|\mathcal{S}_H) \\
    \end{aligned}
\end{eqnarray}
With a pre-specified interim sample size fraction $\omega$, we aim to determine the smallest sample size $n$ and decision boundaries, $\lambda_1$ and $\lambda$, that ensure the PCS under both scenarios of interest achieve pre-specified desired levels, i.e.,  $\beta_L(\lambda_1, \lambda) \geq \alpha_L$ and $\beta_H(\lambda_1, \lambda) \geq \alpha_H$.  To achieve this aim, our strategy is to first identify $\lambda_1$ and $\lambda$ that satisfy $\beta_L(\lambda_1, \lambda) \geq \alpha_L$, based on O'Brien-Fleming error spending function \citep{O'BrienFleming1979}. Given that, we then find the smallest $n$ that satisfies $\beta_H(\lambda_1, \lambda) \geq \alpha_H$.

We first discuss how to identify $\lambda_1$ and $\lambda$ that satisfy $\beta_L(\lambda_1, \lambda) \geq \alpha_L$. This is equivalent to control Pr(select $d_H | \mathcal{S}_L) \leq 1-\alpha_L$, which can be written as 
 \begin{equation} \label{eqn:error_rate_H0}
 \Pr(\widehat{p}_{1,H} - \widehat{p}_{1,L} > \lambda_1 \mid \mathcal{S}_L) + \Pr(\widehat{p}_{1,H} - \widehat{p}_{1,L} \leq \lambda_1, ~  \widehat{p}_{H} - \widehat{p}_{L} > \lambda \mid \mathcal{S}_L) \leq 1 - \alpha_L.
\end{equation}
Let $\sigma_L^2 = 2p_H(1-p_H)$, and define $\lambda^*_1 = \frac{\lambda_1}{\sigma_L/\sqrt{n_1}}$ and $\lambda^* = \frac{\lambda}{\sigma_L/\sqrt{n}}$. The following theorem provides the sampling distribution of (\ref{eqn:error_rate_H0}) and the solution of $\lambda^*_1$ and $\lambda^*$. The proof is provided in Supplementary Material \ref{Sup: proof-theorem2-boundary}. 
\begin{thm} \label{proof-theorem2-boundary}
Define the standardized statistics as $Z_1 = \frac{\widehat{p}_{1,H} -\widehat{p}_{1,L}}{\sigma_L /\sqrt{n_1}}$ for the interim analysis and $Z = \frac{\widehat{p}_{H} -\widehat{p}_{L}}{\sigma_L /\sqrt{n}}$ for the final analysis under $\mathcal{S}_L$. The probability of incorrectly selecting $d_H$ under $\mathcal{S}_L$ can be expressed as:
 \begin{equation} \label{eqn:error_rate_standaerdized_H0}
 \Pr\left(Z_1 > \lambda_1^*\mid \mathcal{S}_L\right) + \Pr\left(Z_1 \leq \lambda_1^*, Z > \lambda^* \mid  \mathcal{S}_L\right) \leq 1 - \alpha_L,
\end{equation}
where $Z_1$ and $Z$ follows a standardized bivariate normal distribution with correlation $\rho = \sqrt{\frac{n_1}{n}} =\sqrt{\omega}$. The value of $\lambda^*_1$ satisfying (\ref{eqn:error_rate_standaerdized_H0}) is given by
$$\lambda^*_1 =\Phi^{-1}\left\{1- 2\Phi\left( \frac{\Phi^{-1}((1-\alpha_L)/2)}{\sqrt{\omega}}\right) \right\},$$
and  $\lambda^*$ is determined by solving equation  (\ref{eqn:error_rate_standaerdized_H0}).
\end{thm}

Given the thresholds $\lambda^*_1$ and $\lambda^*$, we can numerically search for the smallest $n$ that satisfies $\beta_H(\lambda_1, \lambda) \geq \alpha_H$. Detailed fomulas are provided in Supplementary Material \ref{Sec: search_n_two_stage_ROSE}. Once total sample size per dose $n$ is determined, the sample size for the first stage, $n_1$ is obtained by $\lceil \omega n \rceil$, the nearest integer that $\geq \omega n $. The decision boundaries $\lambda_1$ and $\lambda$, can be calculated as  $\lambda_1 = \lambda^*_1\sigma_L/\sqrt{n_1}$ and  $\lambda = \lambda^*\sigma_L/\sqrt{n}$. Supplementary Material \ref{sec:allocarion_ratio_C} describes a generalized two-stage ROSE framework that allows for a flexible allocation ratio $C:1$ between $d_H$ and $d_L$.

Tables \ref{tab:Bounderies_sample_size_delta_0.1} and \ref{tab:Bounderies_sample_size_delta_0.15} present the decision boundaries and minimum sample sizes per dose arm for the two-stage ROSE design, with an interim analysis occurring when 50\% of patients are enrolled and evaluated (i.e., $\omega = 0.5$).  In most cases, the minimum sample size required for the two-stage ROSE design to achieve the same selection accuracy is slightly larger than that of the one-stage design. For example, under the desired accuracy levels of $\alpha_L = \alpha_H = 0.65$, with $\delta = 0.1$ and $p_H = 0.4$, the one-stage design requires a minimum of 28 patients per dose arm. In contrast, the two-stage design requires 31 patients per arm to meet the same accuracy criteria. However, it is important to note that the ‘minimum sample size’ refers to the planned sample size for designing the study; in practice, the actual sample size of the two-stage design is often smaller than that of the one-stage design due to interim stopping (see simulation results later). In addition, both the interim and final decision boundaries in the two-stage design are larger than the boundary used in the one-stage design under the same settings, with the interim boundary $\lambda_1$ consistently exceeding the final boundary $\lambda$.

\subsection{Exact ROSE Design}\label{sec:E-ROSE}
The method described thus far relies on the normal approximation to the binomial distribution, which is widely used in practice for hypothesis testing and sample size determination for binary endpoints. However, due to the approximation, the resulting decision boundary and optimal sample size may not be exact, and the true PCS may slightly deviate from the nominal levels $\alpha_L$ and $\alpha_H$, as described in the simulation section later. In this section, we present the exact ROSE (eROSE) design, in which the decision boundary and sample size are determined using the binomial distribution. In a trial, denote the probability of selecting $d_L$ as $\pi_L$, and the probability of selecting $d_H$ as $\pi_H$, where $\pi_L + \pi_H = 1$. 

Let $k_j$ denotes the number of responses observed in $d_j$, then the $\pi_H$ of one-stage eROSE is calculated as 
\begin{eqnarray} \label{eqn:binom_H1} \nonumber
    \begin{aligned}
      \pi_H = & \sum_{k_L=0}^n\sum_{k_H=0}^{n} I\left[\frac{(k_H-k_L)}{n} > \lambda  \right]
            \times b(k_H; n, p_H) \times  b(k_L; n, p_L) \\
    \end{aligned}
\end{eqnarray}
where $b(k; n, p)$ denotes the binomial probability mass function. 
The exact solution can be obtained by numerically searching for the smallest $n$ and $\lambda$ that satisfy the constraints: 
\[
\beta_L(\lambda) = 1 - \pi_H(\mathcal{S}_L, \lambda) \geq \alpha_L \, \text{ and } \,
\beta_H(\lambda) = \pi_H(\mathcal{S}_H, \lambda) \geq \alpha_H.
\]
The decision boundary $\lambda$ is searched over the range $[0, \delta]$ using a small step size (e.g., 0.01 or 0.001). Due to the discreteness of the binomial distribution, $\lambda$ is not unique. Among all candidate solutions that meet the design criteria, we report the smallest $\lambda$ associated with the minimum sample size $n$ under the specified search granularity. 

In two-stage eROSE design, let  $k_{1,j}$ and $ k_{2,j}$  denote the number of responses observed at stage 1 and stage 2, respectively, for $d_j$. The stage 1 sample size is $n_1 = \lceil \omega n \rceil $, and stage 2 sample size is $n_2 = n - n_1$. The probability of selecting $d_L$ is given by
\begin{eqnarray} \label{eqn:binom_interim_H1} \nonumber
    \begin{aligned}
    \pi_L &= \sum_{k_{1,L}=0}^{n_1}\sum_{k_{1,H}=0}^{n_1} \sum_{k_{2,L}=0}^{n_2}\sum_{k_{2,H}=0}^{n_2} I\left[\frac{(k_{1,H}-k_{1,L})}{n_1} \leq \lambda_1, \frac{(k_{1,H}+k_{2,H}-k_{1,L}-k_{2,L})}{n_1+n_2} \leq \lambda  \right] \\
     & \quad \times b(k_{1,H}; n_{1}, p_{H} ) \times  b(k_{1,L}; n_1, p_L) \times b(k_{2,H}; {n_2}, p_H) \times  b(k_{2,L}; {n_2}, p_L). 
    \end{aligned}
\end{eqnarray}

To align with the two-stage ROSE design, we control the probability of incorrect
interim early termination (i.e., incorrect selection of $d_H$ at the interim) under $\mathcal{S}_L$ using the O’Brien–Fleming
error–spending function. Let $\mathrm{PET}_{S_L}(\lambda_1)$ denote the early
termination probability at the interim under $\mathcal{S}_L$, calculated as $\text{PET}_{S_L}(\lambda_1) = \sum_{k_{1,L}=0}^{n_1}\sum_{k_{1,H}=0}^{n_1} I\left[\frac{k_{1,H}-k_{1,L}}{n_1} > \lambda_1\right]$ $ b(k_{1,H}; n_1, p_H ) b(k_{1,L}; n_1, p_H).$ Specifically, we numerically search for the smallest sample size $n$ and the decision boundaries $\lambda_1$ and $\lambda$ that satisfy the following constraints:
\begin{align*}
\beta_L(\lambda_1, \lambda) &= \pi_L(\mathcal{S}_L; \lambda_1, \lambda) \;\geq\; \alpha_L, \\
\beta_H(\lambda_1, \lambda) &= 1 - \pi_L(\mathcal{S}_H; \lambda_1, \lambda) \;\geq\; \alpha_H, \\
\lambda_1 = \arg\min_{\tilde{\lambda}_1 \in [0,1]} \ \tilde{\lambda}_1 
& \quad \text{s.t.} \quad \mathrm{PET}_{S_L}(\tilde{\lambda}_1) \le 2\Phi\left(\frac{\Phi^{-1}((1-\alpha_L)/2)}{\sqrt{\omega}}\right).
\end{align*}
The interim boundary $\lambda_1$ is chosen as the smallest value in $[0,1]$ that satisfies the incorrect early termination constraint, while the final boundary $\lambda$ is searched over the range $[0, \delta]$. Both searches should use a small step size (e.g., 0.01 or 0.001). 

Tables S7 and S8 in Supplementary Material \ref{sec:eROSE_boundary} summarize the decision boundaries and minimum sample sizes per dose for both one-stage and two-stage eROSE designs using a search granularity of 0.002. Compared with the ROSE design under the same settings, the eROSE design generally requires a similar or larger sample size. Across all settings examined, the differences ranging from 1 patient fewer to 13 patients more per dose arm. Additionally, due to the discreteness of the binomial distribution, the two-stage eROSE design may sometimes require fewer patients than the one-stage version. For instance, when $\alpha_L = \alpha_H = 0.6$, $p_H = 0.3$, and $\delta = 0.1$, the required sample size per dose arm is 23 for the one-stage eROSE design, whereas the two-stage eROSE design requires only 19.

In addition, we provide decision boundaries obtained from the binomial distribution without requiring $\lambda_1$ and $ \lambda$ to follow the O’Brien–Fleming error–spending function. This approach enables a more flexible interim decision rule, subject only to $ \lambda \leq \lambda_1$, and yields the corresponding globally minimum sample size. The results are summarized in Tables S11 and S12.

\section{Simulation Studies}  \label{sec:simulation}
\subsection{Simulation Settings}
To evaluate the operating characteristics of the proposed optimal selection design for each parameter setting, we conducted 10,000 simulations for each parameter setting. The PCS was calculated under both $\mathcal{S}_L$ and $\mathcal{S}_H$, given the decision boundary ($\lambda$) and minimum sample size ($n$) per dose arm listed in Tables \ref{tab:Bounderies_sample_size_delta_0.1} and \ref{tab:Bounderies_sample_size_delta_0.15}. We compared the ROSE design to the U-MET design \citep{DAngelo2024} to assess the PCS of the U-MET design when the response rates of doses and the sample size were matched to those of the ROSE design. Detailed parameter settings for the U-MET design are provided in Supplementary Material \ref{supsec:U-MET}. It is worth noting that the comparison is not intended to demonstrate one design is superior to the other, as the two designs used different criteria to define and select the OBD. Instead, the comparison provides a reference for better understanding the operating characteristics of the ROSE design.

We chose the U-MET design as the comparator because it shares the same objective as our approach, i.e., identifying the OBD among pre-specified candidate doses using data solely from a randomized dose optimization trial. In contrast, the other designs discussed in the Introduction differ in either their objectives or structural frameworks, making direct comparisons less meaningful.  For example,  \citep{Hoering2011} aims to select the dose with the highest efficacy among those deemed tolerable without accounting for benefit-risk trade-offs; U-BOIN \citep{UBOIN2019} and DROID \citep{DROID2023} incorporate data from initial candidate doses determination stage into their randomized phase; and MERIT \citep{MERIT2024} focuses on identifying a set of OBD-admissible doses across multiple endpoints rather than directly evaluating the probability of correctly selecting the single optimal dose.  

\subsection{Simulation Results}
As shown in Table \ref{tab:PCS_ROSE_U-MET}, in most cases, both the PCS when $p_L = p_H$ and the PCS when $p_L = p_H -\delta$ of the ROSE design are close to the pre-specified desired levels $\alpha_L$ and $\alpha_H$.  In some cases, PCS may not precisely reach its respective target level. For example, when the desired accuracy levels are $\alpha_L =\alpha_H= 0.65$, given $p_H = 0.3$ and $\delta = 0.15$, the PCS of selecting $d_H$ when $p_L=0.15 $ is 0.71, while the PCS of selecting $d_L$ when $p_L=p_H =0.3$ is 0.60. This discrepancy arises  from the use of the normal approximation to the binomial distribution, which may be insufficiently accurate, particularly when the sample size is small. Given the early-phase exploratory nature of the randomized dose optimization trials, such minor discrepancy is often ignorable in practice.  Additionally, in practice, the final selection of OBD will also account for other data (e.g., PK/PD, tolerability), in addition to efficacy. The uncertainty due to these factors often outsizes the minor discrepancies caused by discreteness. In trial settings that require strict adherence to pre-specified PCS targets, the eROSE design offers a more suitable option. As demonstrated in Tables S9 and S10, the eROSE design consistently controls PCS at or above the desired levels.

One important feature of the ROSE design, by its construction, is that it allows users to prespecify the target PCS based on clinical and other considerations, thereby ensuring desired operating characteristics. In contrast, the U-MET design does not offer such control, demonstrating highly imbalanced PCS for $d_L$ and $d_H$. Specifically, the U-MET design often exhibits high accuracy in selecting the low dose when $p_L = p_H$, yet it demonstrates poor accuracy in selecting the high dose when $p_L = p_H - \delta$. For example, with a sample size of 24 per dose arm (corresponding to the scenario with $p_H=0.3$, $\delta=0.1$, $\alpha_L=0.6$, and $\alpha_H=0.7$ in Table  \ref{tab:PCS_ROSE_U-MET}), the U-MET design has a PCS of 0.89 for selecting $d_L$, but only 0.27 for selecting $d_H$.

We also evaluated the operating characteristics of the two-stage ROSE design under various parameter configurations, as presented in Table \ref{tab:sim_result_design_interim}.  For each pair of desired selection levels ($\alpha_L$ and $\alpha_H$), we report the PCS, the probability of early termination (PET) in the interim analysis and the expected sample size (EN) per dose arm, under two scenarios: when $p_L=p_H$, i.e., $d_L$ is the OBD, and when $p_L=p_H - \delta$, i.e., $d_H$ is the OBD. The EN of the two-stage design is smaller than the planned sample size due to the possibility of early termination. When $d_L$ is the OBD, the PET is generally low, and the EN remains slightly below but close to the planned sample size. In contrast, when $d_H$ is the OBD, early stopping is more likely, resulting in a reduced EN. For example, when $p_H = 0.3$, $\delta = 0.1$, $\alpha_L =\alpha_H = 0.7$, the two-stage design requires a minimum of 48 patients per arm, but achieves an average sample size of only 39.4 when an interim analysis is conducted after 24 patients. The PCS of the two-stage design reaches the pre-specified selection levels in most cases. Similar to one-stage design, due to the discreteness of the binomial distribution, the PCS may fall slightly below the target in cases with small sample sizes, but the difference is typically ignorable in practice. If strict control of PCSs at the target levels is required, the two-stage eROSE design can be applied.

\subsection{Sensitivity Analysis}
We conducted a sensitivity analysis to evaluate the operating characteristics of the ROSE design under misspecification of $p_H$.  Figure \ref{fig:pH-deviate_0.3} in Supplementary Material \ref{sec:sens_p_deviate} presents the PCSs when the design assumes $p_H = 0.3$, while the true value, denoted by $p_H^{\text{true}}$, deviates from this assumption. The results indicate that the ROSE design is robust to misspecification of $p_H$; the PCS remains close to the desired target levels even when $p_H$ is misspecified.

We also evaluated the operating characteristics of the ROSE design when the actual sample size enrolled in each dose arm deviates from the planned sample size $n$, a common occurrence in practice due to patient dropout, inevaluability, or over-enrollment. The results show that the design is robust to moderate deviations from the planned sample size, with the PCS remaining close to the target levels under such conditions (see Supplementary Material \ref{sec:sens_n_deviate}).

\section{Discussion}  \label{sec:Discussion}
We have proposed an optimal selection design, the ROSE design, to identify the OBD for randomized dose optimization trials. The ROSE design minimizes the sample size while ensuring the PCS at prespecified accuracy levels. One prominent advantage of the ROSE design is its simplicity of implementation. It only involves a simple comparison of the difference in response rates between two dose arms against a predetermined decision boundary. A sample size of 15 to 40 patients per dosage arm typically results in a PCS of 60\% to 70\%. A user-friendly software for implementing ROSE designs is available at www.trialdesign.org.

The two-stage ROSE design can potentially further reduce the sample size when response data can be collected in a timely manner to allow for interim analysis. Although the two-stage design can provide additional sample size savings, whether and how likely that occurs depends on the underlying truth. Given that dose optimization is not solely about comparing response rates but also involves evaluating PK/PD, tolerability, and other data, a one-stage design may be preferred in many cases. This is because it is logistically simpler and allows for more comprehensive data collection, enabling more accurate decision-making based on the totality of the data.

The ROSE design employs a normal approximation, which greatly simplifies the derivation and optimization of the design. Compared to the exact ROSE design, which is based on the exact binomial distribution, the ROSE design may not strictly control the PCS at the nominal levels. However, as previously discussed, given the exploratory nature of early-phase randomized dose optimization trials, such minor discrepancies are typically negligible in practice. Moreover, in real-world settings, the final selection of the OBD also incorporates PK/PD and tolerability data, and the uncertainty associated with these factors often outweighs the small inaccuracies introduced by the approximation. Therefore, the ROSE design is generally sufficient from a practical standpoint. When strict control of the PCS is required, the eROSE design provides a suitable alternative. 

One potential concern with the ROSE design is that it focuses primarily on an efficacy endpoint (with toxicity considered implicitly). In contrast,  dose optimization and the identification of the OBD in practice is a highly complex process that requires a comprehensive evaluation of multiple dimensions, including safety, efficacy, PK/PD, and tolerability—each involving multiple endpoints. Like most statistical methods, the ROSE design aims to offer a simplified yet rigorous framework that captures some key considerations. Given the inherent multi-dimensional complexity, it is virtually impossible to define precise criteria for OBD selection that fully account for all relevant factors. Therefore, the final OBD decision should be guided by both the design’s recommendation and the totality of available data. This can be incorporated into the trial protocol by including appropriate language to allow for practical considerations and flexibility—for example, stating that ``the final OBD selection will be based on the design recommendation and the totality of risk and benefit data".

A closely related potential question is whether a PCS of 60\% or 70\% is sufficiently high when there are two candidate doses. While these values may appear modest, achieving a higher PCS typically requires a larger sample size, often beyond the scope of most early-phase trials. For context, state-of-the-art dose-finding designs generally achieve a PCS of 40\% to 60\% for identifying the MTD \citep{Zhou18CCR}. Moreover, as noted earlier, final dose selection in practice often incorporates additional information—such as PK/PD data and dose-response modeling—especially when the efficacy endpoint does not clearly differentiate between doses.  As a result, the actual PCS can be substantially higher. We do not formally account for all such considerations into the statistical decision framework, as doing so would make the design prohibitively complex and unduly restrictive.

We have considered both one- and two-stage ROSE designs. When appropriate, the design can be extended to multiple stages. In such cases, the decision boundary will not have a closed-form solution. A numerical search is required to identify the decision boundary and the minimal sample size needed to achieve the prespecified PCS levels. In addition, the ROSE design does not include formal toxicity monitoring rule because the doses selected for randomized dose optimization have been previously assessed in phase I trials and should be acceptable in terms of severe toxicities. However, if needed, toxicity monitoring can be easily added to the ROSE design. For example, at the interim and final analysis,  if Pr(toxicity rate $> \phi | data) > C$ in a dose arm, then stop that arm, where $\phi$ is the high limit of the toxicity rate and $C$ is a sample-size-dependent probability cutoff, using the Bayesian optimal phase 2 (BOP2) design \citep{Zhou17SIM}.

\clearpage
\bibliographystyle{plainnat}


\newpage

\begin{table}[ht]
\centering
\caption{Decision boundaries and sample sizes per dose arm for ROSE designs when $\delta=0.1$.}
\label{tab:Bounderies_sample_size_delta_0.1}
\setlength{\tabcolsep}{10pt}
\begin{tabular}{ccccc@{}ccccc}
\hline
\multirow{2}{*}{$p_H$} & \multirow{2}{*}{{$\alpha_L$}} & \multirow{2}{*}{{$\alpha_H$}} & \multicolumn{2}{c}{No interim} &  & \multicolumn{4}{c}{With one interim$^*$} \\ \cline{4-5} \cline{7-10} 
 &  &  & $\lambda$ & $n$ &  & $\lambda_1$ & $n_1$ & $\lambda$ & $n$ \\ \hline
\multirow{10}{*}{0.3} & \multirow{2}{*}{0.60} & 0.60 & 0.052 & 11 &  & 0.178 & 7 & 0.074 & 13 \\ \cline{3-10} 
 &  & 0.70 & 0.034 & 24 &  & 0.126 & 14 & 0.052 & 27 \\ \cline{2-10} 
 & \multirow{2}{*}{0.65} & 0.65 & 0.052 & 24 &  & 0.154 & 14 & 0.065 & 27 \\ \cline{3-10} 
 &  & 0.75 & 0.038 & 44 &  & 0.116 & 25 & 0.048 & 49 \\ \cline{2-10} 
 & \multirow{2}{*}{0.70} & 0.70 & 0.052 & 44 &  & 0.141 & 24 & 0.060 & 48 \\ \cline{3-10} 
 &  & 0.80 & 0.040 & 73 &  & 0.109 & 40 & 0.047 & 79 \\ \cline{2-10} 
 & \multirow{2}{*}{0.75} & 0.75 & 0.052 & 72 &  & 0.131 & 39 & 0.057 & 77 \\ \cline{3-10} 
 &  & 0.85 & 0.041 & 114 &  & 0.105 & 61 & 0.045 & 121 \\ \cline{2-10} 
 & \multirow{2}{*}{0.80} & 0.80 & 0.052 & 112 &  & 0.125 & 59 & 0.055 & 118 \\ \cline{3-10} 
 &  & 0.90 & 0.041 & 176 &  & 0.100 & 92 & 0.044 & 184 \\ \hline
\multirow{10}{*}{0.4} & \multirow{2}{*}{0.60} & 0.60 & 0.051 & 12 &  & 0.190 & 7 & 0.077 & 14 \\ \cline{3-10} 
 &  & 0.70 & 0.033 & 28 &  & 0.126 & 16 & 0.051 & 32 \\ \cline{2-10} 
 & \multirow{2}{*}{0.65} & 0.65 & 0.051 & 28 &  & 0.154 & 16 & 0.065 & 31 \\ \cline{3-10} 
 &  & 0.75 & 0.037 & 52 &  & 0.115 & 29 & 0.048 & 57 \\ \cline{2-10} 
 & \multirow{2}{*}{0.70} & 0.70 & 0.051 & 52 &  & 0.140 & 28 & 0.059 & 56 \\ \cline{3-10} 
 &  & 0.80 & 0.039 & 87 &  & 0.108 & 47 & 0.046 & 93 \\ \cline{2-10} 
 & \multirow{2}{*}{0.75} & 0.75 & 0.051 & 85 &  & 0.130 & 45 & 0.056 & 90 \\ \cline{3-10} 
 &  & 0.85 & 0.040 & 136 &  & 0.103 & 72 & 0.044 & 143 \\ \cline{2-10} 
 & \multirow{2}{*}{0.80} & 0.80 & 0.051 & 132 &  & 0.123 & 69 & 0.054 & 138 \\ \cline{3-10} 
 &  & 0.90 & 0.040 & 209 &  & 0.098 & 109 & 0.043 & 217 \\ \hline
\multirow{10}{*}{0.5} & \multirow{2}{*}{0.60} & 0.60 & 0.050 & 13 &  & 0.181 & 8 & 0.076 & 15 \\ \cline{3-10} 
 &  & 0.70 & 0.033 & 30 &  & 0.124 & 17 & 0.050 & 34 \\ \cline{2-10} 
 & \multirow{2}{*}{0.65} & 0.65 & 0.050 & 30 &  & 0.153 & 17 & 0.064 & 33 \\ \cline{3-10} 
 &  & 0.75 & 0.037 & 56 &  & 0.113 & 31 & 0.047 & 61 \\ \cline{2-10} 
 & \multirow{2}{*}{0.70} & 0.70 & 0.050 & 55 &  & 0.138 & 30 & 0.059 & 59 \\ \cline{3-10} 
 &  & 0.80 & 0.039 & 93 &  & 0.107 & 50 & 0.045 & 99 \\ \cline{2-10} 
 & \multirow{2}{*}{0.75} & 0.75 & 0.050 & 91 &  & 0.129 & 48 & 0.055 & 96 \\ \cline{3-10} 
 &  & 0.85 & 0.040 & 145 &  & 0.102 & 76 & 0.044 & 152 \\ \cline{2-10} 
 & \multirow{2}{*}{0.80} & 0.80 & 0.050 & 141 &  & 0.122 & 73 & 0.053 & 146 \\ \cline{3-10} 
 &  & 0.90 & 0.040 & 223 &  & 0.097 & 116 & 0.043 & 231 \\ \hline
\end{tabular}
\vspace{-0.2em}
\begin{tablenotes}
      \small
      \item  \hspace{30pt} * When the fraction of information $\omega$ reaches 0.5.
\end{tablenotes}
\end{table}

\begin{table}[ht]
\centering
\caption{Decision boundaries and sample sizes for ROSE design when $\delta=0.15$.}
\label{tab:Bounderies_sample_size_delta_0.15}
\setlength{\tabcolsep}{10pt}
\begin{tabular}{ccccc@{}ccccc}
\hline
\multirow{2}{*}{$p_H$} & \multirow{2}{*}{{$\alpha_L$}} & \multirow{2}{*}{{$\alpha_H$}} & \multicolumn{2}{c}{No interim} &  & \multicolumn{4}{c}{With one interim$^*$} \\ \cline{4-5} \cline{7-10} 
 &  &  & $\lambda$ & $n$ &  & $\lambda_1$ & $n_1$ & $\lambda$ & $n$ \\ \hline
\multirow{10}{*}{0.3} & \multirow{2}{*}{0.60} & 0.60 & 0.079 & 5 &  & 0.272 & 3 & 0.110 & 6 \\ \cline{3-10} 
 &  & 0.70 & 0.053 & 10 &  & 0.192 & 6 & 0.077 & 12 \\ \cline{2-10} 
 & \multirow{2}{*}{0.65} & 0.65 & 0.079 & 10 &  & 0.236 & 6 & 0.098 & 12 \\ \cline{3-10} 
 &  & 0.75 & 0.058 & 19 &  & 0.174 & 11 & 0.074 & 21 \\ \cline{2-10} 
 & \multirow{2}{*}{0.70} & 0.70 & 0.079 & 19 &  & 0.209 & 11 & 0.090 & 21 \\ \cline{3-10} 
 &  & 0.80 & 0.062 & 31 &  & 0.168 & 17 & 0.071 & 34 \\ \cline{2-10} 
 & \multirow{2}{*}{0.75} & 0.75 & 0.079 & 31 &  & 0.198 & 17 & 0.085 & 34 \\ \cline{3-10} 
 &  & 0.85 & 0.063 & 49 &  & 0.160 & 26 & 0.069 & 52 \\ \cline{2-10} 
 & \multirow{2}{*}{0.80} & 0.80 & 0.079 & 48 &  & 0.188 & 26 & 0.083 & 51 \\ \cline{3-10} 
 &  & 0.90 & 0.063 & 74 &  & 0.153 & 39 & 0.067 & 78 \\ \hline
\multirow{10}{*}{0.4} & \multirow{2}{*}{0.60} & 0.60 & 0.077 & 6 &  & 0.251 & 4 & 0.108 & 7 \\ \cline{3-10} 
 &  & 0.70 & 0.051 & 12 &  & 0.190 & 7 & 0.077 & 14 \\ \cline{2-10} 
 & \multirow{2}{*}{0.65} & 0.65 & 0.077 & 12 &  & 0.234 & 7 & 0.097 & 14 \\ \cline{3-10} 
 &  & 0.75 & 0.057 & 23 &  & 0.171 & 13 & 0.072 & 25 \\ \cline{2-10} 
 & \multirow{2}{*}{0.70} & 0.70 & 0.077 & 23 &  & 0.205 & 13 & 0.088 & 25 \\ \cline{3-10} 
 &  & 0.80 & 0.060 & 38 &  & 0.161 & 21 & 0.069 & 41 \\ \cline{2-10} 
 & \multirow{2}{*}{0.75} & 0.75 & 0.077 & 37 &  & 0.195 & 20 & 0.084 & 40 \\ \cline{3-10} 
 &  & 0.85 & 0.061 & 59 &  & 0.157 & 31 & 0.068 & 62 \\ \cline{2-10} 
 & \multirow{2}{*}{0.80} & 0.80 & 0.077 & 58 &  & 0.187 & 30 & 0.082 & 60 \\ \cline{3-10} 
 &  & 0.90 & 0.062 & 90 &  & 0.149 & 47 & 0.065 & 94 \\ \hline
\multirow{10}{*}{0.5} & \multirow{2}{*}{0.60} & 0.60 & 0.076 & 6 &  & 0.257 & 4 & 0.111 & 7 \\ \cline{3-10} 
 &  & 0.70 & 0.050 & 14 &  & 0.181 & 8 & 0.076 & 15 \\ \cline{2-10} 
 & \multirow{2}{*}{0.65} & 0.65 & 0.076 & 13 &  & 0.223 & 8 & 0.095 & 15 \\ \cline{3-10} 
 &  & 0.75 & 0.055 & 25 &  & 0.169 & 14 & 0.071 & 27 \\ \cline{2-10} 
 & \multirow{2}{*}{0.70} & 0.70 & 0.076 & 24 &  & 0.209 & 13 & 0.089 & 26 \\ \cline{3-10} 
 &  & 0.80 & 0.058 & 41 &  & 0.161 & 22 & 0.068 & 44 \\ \cline{2-10} 
 & \multirow{2}{*}{0.75} & 0.75 & 0.076 & 40 &  & 0.194 & 21 & 0.084 & 42 \\ \cline{3-10} 
 &  & 0.85 & 0.060 & 64 &  & 0.153 & 34 & 0.066 & 67 \\ \cline{2-10} 
 & \multirow{2}{*}{0.80} & 0.80 & 0.076 & 62 &  & 0.182 & 33 & 0.080 & 65 \\ \cline{3-10} 
 &  & 0.90 & 0.060 & 98 &  & 0.146 & 51 & 0.064 & 102 \\ \hline
\end{tabular}
\vspace{-0.2em}
\begin{tablenotes}
      \small
      \item  \hspace{30pt} * When the fraction of information $\omega$ reaches 0.5.
\end{tablenotes}

\end{table}

\begin{table}[ht]
\centering
\caption{PCS under ROSE and U-MET designs. }
\label{tab:PCS_ROSE_U-MET}
\setlength{\tabcolsep}{5.5pt}
\begin{tabular}{cccccccc@{}cccc}
\hline
\multirow{3}{*}{$p_H$} & \multirow{3}{*}{{$\alpha_L$}} & \multirow{3}{*}{{$\alpha_H$}} & \multicolumn{4}{c}{$\delta$ = 0.1} &  & \multicolumn{4}{c}{$\delta$ = 0.15} \\ \cline{4-7} \cline{9-12} 
 &  &  & \multicolumn{2}{c}{$p_L$= 0.3} & \multicolumn{2}{c}{$p_L$= 0.2} &  & \multicolumn{2}{c}{$p_L$= 0.3} & \multicolumn{2}{c}{$p_L$= 0.15} \\ \cline{4-7} \cline{9-12}
 &  &  & ROSE & U-MET & ROSE & U-MET &  & ROSE & U-MET & ROSE & U-MET \\ \hline
\multirow{10}{*}{0.3} & \multirow{2}{*}{0.60} & 0.60 & 0.59 & 0.89 & 0.62 & 0.20 &  & 0.63 & 0.87 & 0.58 & 0.23 \\ \cline{3-12} 
 &  & 0.70 & 0.57 & 0.89 & 0.74 & 0.27 &  & 0.59 & 0.88 & 0.71 & 0.28 \\ \cline{2-12} 
 & \multirow{2}{*}{0.65} & 0.65 & 0.68 & 0.89 & 0.62 & 0.27 &  & 0.60 & 0.88 & 0.71 & 0.28 \\ \cline{3-12} 
 &  & 0.75 & 0.64 & 0.90 & 0.75 & 0.33 &  & 0.70 & 0.89 & 0.71 & 0.36 \\ \cline{2-12} 
 & \multirow{2}{*}{0.70} & 0.70 & 0.72 & 0.90 & 0.68 & 0.33 &  & 0.70 & 0.89 & 0.70 & 0.36 \\ \cline{3-12} 
 &  & 0.80 & 0.68 & 0.92 & 0.82 & 0.40 &  & 0.65 & 0.90 & 0.83 & 0.44 \\ \cline{2-12} 
 & \multirow{2}{*}{0.75} & 0.75 & 0.73 & 0.92 & 0.76 & 0.40 &  & 0.76 & 0.90 & 0.75 & 0.44 \\ \cline{3-12} 
 &  & 0.85 & 0.74 & 0.93 & 0.86 & 0.49 &  & 0.78 & 0.91 & 0.83 & 0.54 \\ \cline{2-12} 
 & \multirow{2}{*}{0.80} & 0.80 & 0.79 & 0.93 & 0.82 & 0.48 &  & 0.78 & 0.91 & 0.82 & 0.54 \\ \cline{3-12} 
 &  & 0.90 & 0.81 & 0.94 & 0.89 & 0.60 &  & 0.79 & 0.92 & 0.91 & 0.65 \\ \hline
&  &    & \multicolumn{2}{c}{$p_L$= 0.4} & \multicolumn{2}{c}{$p_L$= 0.3} &  & \multicolumn{2}{c}{$p_L$= 0.4} & \multicolumn{2}{c}{$p_L$= 0.25} \\ \cline{4-7} \cline{9-12} 
 &  &  & ROSE & U-MET & ROSE & U-MET &  & ROSE & U-MET & ROSE & U-MET \\ \hline
\multirow{10}{*}{0.4} & \multirow{2}{*}{0.60} & 0.60 & 0.59 & 0.88 & 0.61 & 0.24 &  & 0.61 & 0.85 & 0.61 & 0.27 \\ \cline{3-12} 
 &  & 0.70 & 0.56 & 0.88 & 0.74 & 0.29 &  & 0.58 & 0.88 & 0.71 & 0.31 \\ \cline{2-12} 
 & \multirow{2}{*}{0.65} & 0.65 & 0.66 & 0.88 & 0.65 & 0.29 &  & 0.59 & 0.88 & 0.72 & 0.31 \\ \cline{3-12} 
 &  & 0.75 & 0.62 & 0.90 & 0.78 & 0.36 &  & 0.67 & 0.88 & 0.73 & 0.40 \\ \cline{2-12} 
 & \multirow{2}{*}{0.70} & 0.70 & 0.69 & 0.90 & 0.72 & 0.36 &  & 0.68 & 0.89 & 0.73 & 0.38 \\ \cline{3-12} 
 &  & 0.80 & 0.70 & 0.91 & 0.80 & 0.43 &  & 0.72 & 0.90 & 0.79 & 0.48 \\ \cline{2-12} 
 & \multirow{2}{*}{0.75} & 0.75 & 0.76 & 0.92 & 0.73 & 0.43 &  & 0.73 & 0.89 & 0.77 & 0.46 \\ \cline{3-12} 
 &  & 0.85 & 0.76 & 0.93 & 0.85 & 0.52 &  & 0.74 & 0.91 & 0.86 & 0.57 \\ \cline{2-12} 
 & \multirow{2}{*}{0.80} & 0.80 & 0.79 & 0.93 & 0.81 & 0.52 &  & 0.80 & 0.90 & 0.80 & 0.56 \\ \cline{3-12} 
 &  & 0.90 & 0.80 & 0.94 & 0.90 & 0.63 &  & 0.80 & 0.91 & 0.90 & 0.68 \\ \hline
 &  &  & \multicolumn{2}{c}{$p_L$= 0.5} & \multicolumn{2}{c}{$p_L$= 0.4} &  & \multicolumn{2}{c}{$p_L$= 0.5} & \multicolumn{2}{c}{$p_L$= 0.35} \\ \cline{4-7} \cline{9-12} 
 &  &  & ROSE & U-MET & ROSE & U-MET &  & ROSE & U-MET & ROSE & U-MET \\ \hline
\multirow{10}{*}{0.5} & \multirow{2}{*}{0.60} & 0.60 & 0.58 & 0.86 & 0.62 & 0.26 &  & 0.61 & 0.64 & 0.61 & 0.29 \\ \cline{3-12} 
 &  & 0.70 & 0.55 & 0.88 & 0.74 & 0.30 &  & 0.58 & 0.87 & 0.74 & 0.33 \\ \cline{2-12} 
 & \multirow{2}{*}{0.65} & 0.65 & 0.65 & 0.88 & 0.65 & 0.30 &  & 0.57 & 0.86 & 0.72 & 0.33 \\ \cline{3-12} 
 &  & 0.75 & 0.68 & 0.90 & 0.72 & 0.38 &  & 0.66 & 0.88 & 0.74 & 0.40 \\ \cline{2-12} 
 & \multirow{2}{*}{0.70} & 0.70 & 0.68 & 0.90 & 0.72 & 0.37 &  & 0.67 & 0.87 & 0.74 & 0.40 \\ \cline{3-12} 
 &  & 0.80 & 0.69 & 0.91 & 0.81 & 0.45 &  & 0.71 & 0.89 & 0.79 & 0.49 \\ \cline{2-12} 
 & \multirow{2}{*}{0.75} & 0.75 & 0.75 & 0.91 & 0.75 & 0.44 &  & 0.79 & 0.88 & 0.72 & 0.49 \\ \cline{3-12} 
 &  & 0.85 & 0.73 & 0.92 & 0.86 & 0.53 &  & 0.73 & 0.90 & 0.86 & 0.60 \\ \cline{2-12} 
 & \multirow{2}{*}{0.80} & 0.80 & 0.82 & 0.92 & 0.79 & 0.54 &  & 0.79 & 0.90 & 0.81 & 0.57 \\ \cline{3-12} 
 &  & 0.90 & 0.79 & 0.94 & 0.91 & 0.64 &  & 0.78 & 0.91 & 0.92 & 0.70 \\ \hline
\end{tabular}
\end{table}

\begin{table}[ht]
\centering
\caption{Simulation results of the two-stage ROSE design, with interim monitoring conducted when half of the patients are enrolled and evaluated, i.e., $\omega=0.5$.}
\label{tab:sim_result_design_interim}
 \setlength{\tabcolsep}{4.5pt}
\begin{tabular}{cccccccccc@{}cccccc}
\hline
\multirow{3}{*}{$p_H$} & \multirow{3}{*}{{$\alpha_L$}} & \multirow{3}{*}{{$\alpha_H$}} & \multicolumn{6}{c}{$\delta = 0.1$} &  & \multicolumn{6}{c}{$\delta = 0.15$} \\ \cline{4-9} \cline{11-16}  
&  &  & \multicolumn{3}{c}{$p_L=0.3$} & \multicolumn{3}{c}{$p_L=0.2$} &  & \multicolumn{3}{c}{$p_L=0.3$} & \multicolumn{3}{c}{$p_L=0.15$} \\ \cline{4-9} \cline{11-16} 
 &  &  & PCS & PET & EN & PCS & PET & EN &  & PCS & PET & EN & PCS & PET & EN \\ \hline
\multirow{10}{*}{0.3} & \multirow{2}{*}{0.60} & 0.60 & 0.57 & 0.19 & 11.9 & 0.65 & 0.31 & 11.1 &  & 0.55 & 0.32 & 5.0 & 0.67 & 0.47 & 4.6 \\ \cline{3-16} 
 &  & 0.70 & 0.60 & 0.27 & 23.5 & 0.70 & 0.48 & 20.7 &  & 0.56 & 0.17 & 11.0 & 0.75 & 0.33 & 10.0 \\ \cline{2-16} 
 & \multirow{2}{*}{0.65} & 0.65 & 0.65 & 0.15 & 25.1 & 0.66 & 0.31 & 23.0 &  & 0.69 & 0.17 & 11.0 & 0.59 & 0.33 & 10.0 \\ \cline{3-16} 
 &  & 0.75 & 0.63 & 0.23 & 43.5 & 0.74 & 0.50 & 37.1 &  & 0.62 & 0.24 & 18.6 & 0.77 & 0.53 & 15.7 \\ \cline{2-16} 
 & \multirow{2}{*}{0.70} & 0.70 & 0.68 & 0.14 & 44.7 & 0.72 & 0.36 & 39.4 &  & 0.68 & 0.12 & 19.8 & 0.74 & 0.32 & 17.8 \\ \cline{3-16} 
 &  & 0.80 & 0.70 & 0.13 & 73.9 & 0.81 & 0.44 & 61.7 &  & 0.69 & 0.18 & 31.0 & 0.80 & 0.51 & 25.3 \\ \cline{2-16} 
 & \multirow{2}{*}{0.75} & 0.75 & 0.76 & 0.09 & 73.7 & 0.74 & 0.34 & 63.9 &  & 0.72 & 0.10 & 32.3 & 0.78 & 0.34 & 28.2 \\ \cline{3-16} 
 &  & 0.85 & 0.75 & 0.10 & 114.8 & 0.85 & 0.47 & 92.5 &  & 0.75 & 0.09 & 49.7 & 0.86 & 0.43 & 40.9 \\ \cline{2-16} 
 & \multirow{2}{*}{0.80} & 0.80 & 0.80 & 0.07 & 114.2 & 0.80 & 0.36 & 96.5 &  & 0.81 & 0.09 & 48.9 & 0.79 & 0.42 & 40.5 \\ \cline{3-16} 
 &  & 0.90 & 0.81 & 0.06 & 178.2 & 0.89 & 0.49 & 138.9 &  & 0.80 & 0.09 & 74.5 & 0.89 & 0.54 & 56.8 \\ \hline
\multicolumn{1}{l}{} & \multicolumn{1}{l}{} &  & \multicolumn{3}{c}{$p_L=0.4$} & \multicolumn{3}{c}{$p_L=0.3$} &  & \multicolumn{3}{c}{$p_L=0.4$} & \multicolumn{3}{c}{$p_L=0.25$} \\ \cline{4-9} \cline{11-16} 
\multicolumn{1}{l}{} & \multicolumn{1}{l}{} & \multicolumn{1}{l}{} & PCS & PET & EN & PCS & PET & EN &  & PCS & PET & EN & PCS & PET & EN \\ \hline
\multirow{10}{*}{0.4} & \multirow{2}{*}{0.60} & 0.60 & 0.66 & 0.21 & 12.6 & 0.54 & 0.33 & 11.7 &  & 0.59 & 0.14 & 6.6 & 0.64 & 0.25 & 6.3 \\ \cline{3-16} 
 &  & 0.70 & 0.61 & 0.19 & 29.0 & 0.70 & 0.37 & 26.1 &  & 0.66 & 0.20 & 12.6 & 0.65 & 0.40 & 11.2 \\ \cline{2-16} 
 & \multirow{2}{*}{0.65} & 0.65 & 0.68 & 0.18 & 28.2 & 0.61 & 0.37 & 25.4 &  & 0.66 & 0.21 & 12.6 & 0.64 & 0.39 & 11.3 \\ \cline{3-16} 
 &  & 0.75 & 0.65 & 0.17 & 52.1 & 0.76 & 0.43 & 44.8 &  & 0.63 & 0.16 & 23.0 & 0.77 & 0.41 & 20.1 \\ \cline{2-16} 
 & \multirow{2}{*}{0.70} & 0.70 & 0.70 & 0.16 & 51.5 & 0.71 & 0.42 & 44.2 &  & 0.71 & 0.16 & 23.0 & 0.68 & 0.41 & 20.1 \\ \cline{3-16} 
 &  & 0.80 & 0.72 & 0.12 & 87.4 & 0.79 & 0.43 & 73.0 &  & 0.68 & 0.14 & 38.3 & 0.82 & 0.45 & 32.0 \\ \cline{2-16} 
 & \multirow{2}{*}{0.75} & 0.75 & 0.76 & 0.12 & 84.6 & 0.74 & 0.41 & 71.3 &  & 0.75 & 0.13 & 37.3 & 0.75 & 0.43 & 31.4 \\ \cline{3-16} 
 &  & 0.85 & 0.76 & 0.10 & 135.8 & 0.84 & 0.47 & 109.5 &  & 0.76 & 0.12 & 58.3 & 0.85 & 0.52 & 45.9 \\ \cline{2-16} 
 & \multirow{2}{*}{0.80} & 0.80 & 0.80 & 0.07 & 133.2 & 0.81 & 0.39 & 111.2 &  & 0.78 & 0.07 & 57.8 & 0.82 & 0.39 & 48.3 \\ \cline{3-16} 
 &  & 0.90 & 0.80 & 0.07 & 209.0 & 0.90 & 0.53 & 160.2 &  & 0.82 & 0.06 & 91.3 & 0.89 & 0.46 & 72.4 \\ \hline
 &  &  & \multicolumn{3}{c}{$p_L=0.5$} & \multicolumn{3}{c}{$p_L=0.4$} &  & \multicolumn{3}{c}{$p_L=0.5$} & \multicolumn{3}{c}{$p_L=0.35$} \\ \cline{4-9} \cline{11-16} 
\multicolumn{1}{l}{} & \multicolumn{1}{l}{} & \multicolumn{1}{l}{} & PCS & PET & EN & PCS & PET & EN &  & PCS & PET & EN & PCS & PET & EN \\ \hline
\multirow{10}{*}{0.5} & \multirow{2}{*}{0.60} & 0.60 & 0.64 & 0.22 & 13.4 & 0.56 & 0.37 & 12.4 &  & 0.60 & 0.15 & 6.6 & 0.62 & 0.27 & 6.2 \\ \cline{3-16} 
 &  & 0.70 & 0.60 & 0.20 & 30.7 & 0.71 & 0.39 & 27.3 &  & 0.64 & 0.23 & 13.4 & 0.67 & 0.44 & 11.9 \\ \cline{2-16} 
 & \multirow{2}{*}{0.65} & 0.65 & 0.68 & 0.20 & 29.8 & 0.63 & 0.39 & 26.8 &  & 0.64 & 0.23 & 13.4 & 0.67 & 0.45 & 11.9 \\ \cline{3-16} 
 &  & 0.75 & 0.64 & 0.19 & 55.4 & 0.77 & 0.46 & 47.3 &  & 0.62 & 0.18 & 24.7 & 0.79 & 0.44 & 21.2 \\ \cline{2-16} 
 & \multirow{2}{*}{0.70} & 0.70 & 0.71 & 0.12 & 55.5 & 0.69 & 0.34 & 49.1 &  & 0.70 & 0.16 & 23.9 & 0.69 & 0.42 & 20.6 \\ \cline{3-16} 
 &  & 0.80 & 0.70 & 0.14 & 92.0 & 0.81 & 0.47 & 75.8 &  & 0.67 & 0.15 & 40.6 & 0.84 & 0.49 & 33.3 \\ \cline{2-16} 
 & \multirow{2}{*}{0.75} & 0.75 & 0.77 & 0.09 & 91.8 & 0.74 & 0.36 & 78.9 &  & 0.76 & 0.08 & 40.3 & 0.75 & 0.34 & 34.8 \\ \cline{3-16} 
 &  & 0.85 & 0.74 & 0.12 & 143.1 & 0.86 & 0.51 & 113.5 &  & 0.76 & 0.09 & 64.1 & 0.85 & 0.46 & 51.8 \\ \cline{2-16} 
 & \multirow{2}{*}{0.80} & 0.80 & 0.79 & 0.08 & 140.1 & 0.81 & 0.42 & 115.1 &  & 0.80 & 0.10 & 61.9 & 0.80 & 0.46 & 50.4 \\ \cline{3-16} 
 &  & 0.90 & 0.79 & 0.07 & 223.3 & 0.90 & 0.50 & 174.0 &  & 0.80 & 0.07 & 98.5 & 0.91 & 0.51 & 75.8 \\ \hline
\end{tabular}
\end{table}

\clearpage\pagebreak

 \baselineskip=24pt
 \begin{center}
{\Large \bf Supplementary Materials of ``Randomized Optimal Selection Design for Dose Optimization"}
\end{center}
\begin{center}
{\bf Shuqi Wang, Ying Yuan$^*$, Suyu Liu$^{**}$}
\end{center}

\begin{center}
Department of Biostatistics, The University of Texas MD Anderson Cancer Center\\
Houston, TX, USA.\\
{*Email: yyuan@mdanderson.org;  $\qquad$ **Email: syliu@mdanderson.org}\\
\end{center}

\beginsupplement

\setcounter{equation}{0} \setcounter{table}{0} \setcounter{page}{1}
\setcounter{figure}{0}\setcounter{section}{0}

\section{\texorpdfstring{Approximation of PCS under $\mathcal{S}_L$ and $\mathcal{S}_H$}{Approximation of PCS}} \label{sec:approx_PCS}
As a general rule of thumb, when $np \geq 5$ and $n(1-p)\geq 5$, the sample proportion $\widehat{p}$ can be well approximated by the normal distribution $N(p, \, p(1-p)/n)$ \citep{Fleiss2013}. Consequently, the sampling distribution of $\widehat{p}_H - \widehat{p}_L$ can be approximated by 
$$\widehat{p}_H - \widehat{p}_L \sim N(p_H-p_L,   \frac{p_L(1-p_L) + p_H(1-p_H)}{n}).$$
It follows that
\begin{eqnarray} \label{eqn:PCS_H0_cal} \nonumber
      \beta_L(\lambda)  &=& \Pr(\widehat{p}_H - \widehat{p}_L \leq \lambda \mid \mathcal{S}_L)  \\ \nonumber
           &\approx& \Pr\left(\frac{\sqrt{n}(\widehat{p}_H-\widehat{p}_L)}{\sqrt{2 p_H(1-p_H)}} \leq \frac{\sqrt{n}\lambda}{\sqrt{2 p_H(1-p_H)}}\mid \mathcal{S}_L\right)  \\ \nonumber
           &=& \Phi\left(\frac{\sqrt{n}\lambda}{\sqrt{2 p_H(1-p_H)}} \mid \mathcal{S}_L\right)
\end{eqnarray}
and
\begin{eqnarray} \label{eqn:PCS_H1_cal} \nonumber
      \beta_H(\lambda) &=&  \Pr(\widehat{p}_H - \widehat{p}_L > \lambda \mid \mathcal{S}_H) \\ \nonumber
           & \approx & \Pr\left(\frac{\sqrt{n}(\widehat{p}_H-\widehat{p}_L -\delta)}{\sqrt{(p_H - \delta)(1-p_H+\delta) +p_H(1-p_H)}} > \frac{\sqrt{n}(\lambda-\delta)}{\sqrt{(p_H - \delta)(1-p_H+\delta) +p_H(1-p_H)}}\mid \mathcal{S}_H\right)\\ \nonumber
           & = & 1- \Phi\left(\frac{\sqrt{n}(\lambda-\delta)}{\sqrt{(p_H - \delta)(1-p_H+\delta) +p_H(1-p_H)}}  \mid \mathcal{S}_H \right),
\end{eqnarray}
where $\Phi(\cdot)$ is the cumulative distribution function (CDF) of a $N(0,1)$ distribution. $\beta_H(\lambda)$ is derived under $\mathcal{S}_H$ with $p_L = p_H - \delta$. Given $p_H$ and $\delta$, ensuring adequate PCS under this setting guarantees sufficient PCS for all $p_L < p_H - \delta$.

\section{Proof of Theorem 1}\label{Sup: proof-theorem1-samplesize}

To satisfy the pre-specified requirement of correct selection probabilities, the lower bound of the decision boundary is given by $\lambda_L = \frac{\sqrt{2 p_H(1-p_H)}}{\sqrt{n}}\Phi^{-1}(\alpha_L)$, and the upper bound is $\lambda_H =\frac{\sqrt{(p_H - \delta)(1-p_H+\delta) +p_H(1-p_H)}}{\sqrt{n}}\Phi^{-1}(1-\alpha_H) + \delta$. The desired levels are typically higher than 0.5 since the PCS between two doses is 0.5 under random choice. Under this condition, $\Phi^{-1}(\alpha_L)>0$ and $\Phi^{-1}(1-\alpha_H)< 0$.  Consequently, $\lambda_L$ increases monotonically and $\lambda_H$ decreases monotonically as the sample size $n$ decreases. In other words, as $n$ decreases, the interval $\mathcal{A}_n(\alpha_L, \alpha_H)$ becomes smaller. When the $\mathcal{A}_n(\alpha_L, \alpha_H)$ reduces to a single value, i,e., $\lambda_L = \lambda_H$, $n$ is the smallest value that the correct selection probabilities of $d_H$ and $d_L$. 

\newpage
\section{\texorpdfstring{Decision boundary, sample size per arm, and operating characteristics of the ROSE design when $\delta=0.05$}{Decision boundary and sample size per arm for ROSE design when delta=0.05}} \label{sec:ROSE_delta_0.05}

\begin{table}[H]
\centering
\caption{Decision boundaries and sample sizes per dose arm for ROSE designs when $\delta=0.05$.} 
\label{tab:Bounderies_sample_size_delta_0.05}
\setlength{\tabcolsep}{10pt}
\begin{tabular}{ccccc@{}ccccc}
\hline
\multirow{2}{*}{$p_H$} & \multirow{2}{*}{$\alpha_L$} & \multirow{2}{*}{$\alpha_H$} & \multicolumn{2}{c}{No interim} &  & \multicolumn{4}{c}{With one interim$^*$} \\ \cline{4-5} \cline{7-10} 
 &  &  & $\lambda$ & $n$ &  & $\lambda_1$ & $n_1$ & $\lambda$ & $n$ \\ \hline
\multirow{10}{*}{0.3} & \multirow{2}{*}{0.60} & 0.60 & 0.025 & 42 &  & 0.096 & 24 & 0.039 & 48 \\ \cline{3-10} 
 &  & 0.70 & 0.017 & 98 &  & 0.063 & 55 & 0.026 & 110 \\ \cline{2-10} 
 & \multirow{2}{*}{0.65} & 0.65 & 0.025 & 98 &  & 0.079 & 54 & 0.033 & 108 \\ \cline{3-10} 
 &  & 0.75 & 0.018 & 183 &  & 0.058 & 100 & 0.024 & 200 \\ \cline{2-10} 
 & \multirow{2}{*}{0.70} & 0.70 & 0.025 & 180 &  & 0.070 & 98 & 0.030 & 195 \\ \cline{3-10} 
 &  & 0.80 & 0.020 & 304 &  & 0.054 & 163 & 0.023 & 325 \\ \cline{2-10} 
 & \multirow{2}{*}{0.75} & 0.75 & 0.025 & 298 &  & 0.065 & 158 & 0.028 & 316 \\ \cline{3-10} 
 &  & 0.85 & 0.020 & 476 &  & 0.052 & 251 & 0.022 & 502 \\ \cline{2-10} 
 & \multirow{2}{*}{0.80} & 0.80 & 0.025 & 464 &  & 0.062 & 242 & 0.027 & 484 \\ \cline{3-10} 
 &  & 0.90 & 0.020 & 733 &  & 0.049 & 381 & 0.021 & 761 \\ \hline
\multirow{10}{*}{0.4} & \multirow{2}{*}{0.60} & 0.60 & 0.025 & 49 &  & 0.095 & 28 & 0.039 & 55 \\ \cline{3-10} 
 &  & 0.70 & 0.016 & 115 &  & 0.063 & 64 & 0.025 & 127 \\ \cline{2-10} 
 & \multirow{2}{*}{0.65} & 0.65 & 0.025 & 113 &  & 0.078 & 62 & 0.032 & 124 \\ \cline{3-10} 
 &  & 0.75 & 0.018 & 213 &  & 0.057 & 116 & 0.024 & 231 \\ \cline{2-10} 
 & \multirow{2}{*}{0.70} & 0.70 & 0.025 & 209 &  & 0.070 & 113 & 0.029 & 225 \\ \cline{3-10} 
 &  & 0.80 & 0.019 & 353 &  & 0.054 & 189 & 0.023 & 377 \\ \cline{2-10} 
 & \multirow{2}{*}{0.75} & 0.75 & 0.025 & 345 &  & 0.065 & 183 & 0.028 & 365 \\ \cline{3-10} 
 &  & 0.85 & 0.020 & 554 &  & 0.051 & 291 & 0.022 & 582 \\ \cline{2-10} 
 & \multirow{2}{*}{0.80} & 0.80 & 0.025 & 537 &  & 0.061 & 280 & 0.027 & 559 \\ \cline{3-10} 
 &  & 0.90 & 0.020 & 852 &  & 0.049 & 442 & 0.021 & 883 \\ \hline
\multirow{10}{*}{0.5} & \multirow{2}{*}{0.60} & 0.60 & 0.025 & 52 &  & 0.095 & 29 & 0.038 & 58 \\ \cline{3-10} 
 &  & 0.70 & 0.016 & 121 &  & 0.063 & 67 & 0.025 & 134 \\ \cline{2-10} 
 & \multirow{2}{*}{0.65} & 0.65 & 0.025 & 119 &  & 0.078 & 65 & 0.032 & 130 \\ \cline{3-10} 
 &  & 0.75 & 0.018 & 224 &  & 0.057 & 122 & 0.024 & 243 \\ \cline{2-10} 
 & \multirow{2}{*}{0.70} & 0.70 & 0.025 & 220 &  & 0.070 & 118 & 0.029 & 235 \\ \cline{3-10} 
 &  & 0.80 & 0.019 & 373 &  & 0.054 & 199 & 0.023 & 397 \\ \cline{2-10} 
 & \multirow{2}{*}{0.75} & 0.75 & 0.025 & 364 &  & 0.064 & 192 & 0.028 & 383 \\ \cline{3-10} 
 &  & 0.85 & 0.020 & 584 &  & 0.051 & 307 & 0.022 & 613 \\ \cline{2-10} 
 & \multirow{2}{*}{0.80} & 0.80 & 0.025 & 566 &  & 0.061 & 294 & 0.027 & 588 \\ \cline{3-10} 
 &  & 0.90 & 0.020 & 899 &  & 0.048 & 466 & 0.021 & 931 \\ \hline
\end{tabular}
\vspace{-0.2em}
\begin{tablenotes}
      \small
      \item  \hspace{30pt} * When the fraction of information $\omega$ reaches 0.5.
\end{tablenotes}
\end{table}

\newpage
\begin{table}[H]
\centering
\caption{Simulation results of one-stage and two-stage ROSE design when $\delta = 0.05$.}
\label{tab:sim_result_design_delta_0.05}
\begin{tabular}{cccccccccccc}
\hline
\multirow{3}{*}{$p_H$} & \multirow{3}{*}{$\alpha_L$} & \multirow{3}{*}{$\alpha_H$} & \multicolumn{2}{c}{No interim} &  & \multicolumn{6}{c}{With one interim$^*$} \\ \cline{4-5} \cline{7-12} 
 &  &   & \multicolumn{2}{c}{PCS} &  & \multicolumn{3}{c}{$p_L=0.3$} & \multicolumn{3}{c}{$p_L=0.25$} \\ \cline{4-5} \cline{7-12} 
 &   &  & $p_L=0.3$ & $p_L=0.25$ &  &PCS  &PET  &EN   &PCS  &PET  &EN  \\ \hline
\multirow{10}{*}{0.3} & \multirow{2}{*}{0.60} & 0.60 & 0.64 & 0.55 &  & 0.59 & 0.21 & 42.9 & 0.61 & 0.34 & 39.9 \\ \cline{3-12} 
 &  & 0.70 & 0.59 & 0.71 &  & 0.58 & 0.24 & 97.0 & 0.71 & 0.44 & 85.9 \\ \cline{2-12} 
 & \multirow{2}{*}{0.65} & 0.65 & 0.65 & 0.65 &  & 0.66 & 0.17 & 98.7 & 0.65 & 0.36 & 88.8 \\ \cline{3-12} 
 &  & 0.75 & 0.67 & 0.74 &  & 0.64 & 0.19 & 180.8 & 0.76 & 0.47 & 153.4 \\ \cline{2-12} 
 & \multirow{2}{*}{0.70} & 0.70 & 0.71 & 0.71 &  & 0.68 & 0.16 & 179.4 & 0.71 & 0.40 & 155.8 \\ \cline{3-12} 
 &  & 0.80 & 0.68 & 0.81 &  & 0.69 & 0.15 & 300.7 & 0.81 & 0.48 & 246.5 \\ \cline{2-12} 
 & \multirow{2}{*}{0.75} & 0.75 & 0.74 & 0.75 &  & 0.74 & 0.10 & 300.4 & 0.76 & 0.37 & 257.0 \\ \cline{3-12} 
 &  & 0.85 & 0.75 & 0.85 &  & 0.76 & 0.11 & 473.7 & 0.84 & 0.50 & 376.7 \\ \cline{2-12} 
 & \multirow{2}{*}{0.80} & 0.80 & 0.80 & 0.81 &  & 0.81 & 0.08 & 465.8 & 0.80 & 0.40 & 388.1 \\ \cline{3-12} 
 &  & 0.90 & 0.79 & 0.90 &  & 0.80 & 0.07 & 733.3 & 0.90 & 0.51 & 567.0 \\ \hline
 &  &   & \multicolumn{2}{c}{PCS} &  & \multicolumn{3}{c}{$p_L=0.4$} & \multicolumn{3}{c}{$p_L=0.35$} \\ \cline{4-5} \cline{7-12} 
 &   &  & $p_L=0.4$ & $p_L=0.35$ &  &PCS  &PET  &EN   &PCS  &PET  &EN  \\ \hline 
\multirow{10}{*}{0.4} & \multirow{2}{*}{0.60} & 0.60 & 0.61 & 0.59 &  & 0.60 & 0.26 & 48.1 & 0.58 & 0.38 & 44.8 \\ \cline{3-12} 
 &  & 0.70 & 0.58 & 0.72 &  & 0.63 & 0.21 & 113.8 & 0.67 & 0.40 & 101.9 \\ \cline{2-12} 
 & \multirow{2}{*}{0.65} & 0.65 & 0.63 & 0.67 &  & 0.66 & 0.20 & 111.3 & 0.64 & 0.39 & 99.6 \\ \cline{3-12} 
 &  & 0.75 & 0.64 & 0.76 &  & 0.65 & 0.19 & 208.9 & 0.75 & 0.46 & 178.3 \\ \cline{2-12} 
 & \multirow{2}{*}{0.70} & 0.70 & 0.70 & 0.70 &  & 0.70 & 0.15 & 208.3 & 0.70 & 0.40 & 180.5 \\ \cline{3-12} 
 &  & 0.80 & 0.69 & 0.81 &  & 0.70 & 0.14 & 351.4 & 0.80 & 0.46 & 290.7 \\ \cline{2-12} 
 & \multirow{2}{*}{0.75} & 0.75 & 0.75 & 0.75 &  & 0.75 & 0.11 & 344.1 & 0.74 & 0.39 & 293.8 \\ \cline{3-12} 
 &  & 0.85 & 0.76 & 0.84 &  & 0.75 & 0.11 & 550.6 & 0.86 & 0.50 & 435.9 \\ \cline{2-12} 
 & \multirow{2}{*}{0.80} & 0.80 & 0.80 & 0.81 &  & 0.80 & 0.07 & 540.4 & 0.80 & 0.37 & 455.5 \\ \cline{3-12} 
 &  & 0.90 & 0.80 & 0.89 &  & 0.79 & 0.07 & 852.0 & 0.90 & 0.51 & 658.6 \\ \hline
 &  &   & \multicolumn{2}{c}{PCS} &  & \multicolumn{3}{c}{$p_L=0.5$} & \multicolumn{3}{c}{$p_L=0.45$} \\ \cline{4-5} \cline{7-12} 
 &   &  & $p_L=0.5$ & $p_L=0.45$ &  &PCS  &PET  &EN   &PCS  &PET  &EN  \\ \hline
\multirow{10}{*}{0.5} & \multirow{2}{*}{0.60} & 0.60 & 0.62 & 0.58 &  & 0.60 & 0.26 & 50.6 & 0.60 & 0.39 & 46.6 \\ \cline{3-12} 
 &  & 0.70 & 0.58 & 0.72 &  & 0.61 & 0.22 & 119.3 & 0.70 & 0.42 & 105.6 \\ \cline{2-12} 
 & \multirow{2}{*}{0.65} & 0.65 & 0.63 & 0.67 &  & 0.66 & 0.17 & 119.2 & 0.64 & 0.35 & 107.2 \\ \cline{3-12} 
 &  & 0.75 & 0.67 & 0.73 &  & 0.64 & 0.20 & 218.5 & 0.76 & 0.48 & 185.2 \\ \cline{2-12} 
 & \multirow{2}{*}{0.70} & 0.70 & 0.70 & 0.70 &  & 0.69 & 0.13 & 219.8 & 0.71 & 0.36 & 192.4 \\ \cline{3-12} 
 &  & 0.80 & 0.71 & 0.79 &  & 0.69 & 0.14 & 368.4 & 0.81 & 0.47 & 303.3 \\ \cline{2-12} 
 & \multirow{2}{*}{0.75} & 0.75 & 0.76 & 0.74 &  & 0.76 & 0.09 & 364.9 & 0.75 & 0.38 & 311.1 \\ \cline{3-12} 
 &  & 0.85 & 0.74 & 0.85 &  & 0.76 & 0.10 & 583.2 & 0.84 & 0.50 & 461.0 \\ \cline{2-12} 
 & \multirow{2}{*}{0.80} & 0.80 & 0.80 & 0.79 &  & 0.80 & 0.07 & 567.4 & 0.81 & 0.41 & 467.6 \\ \cline{3-12} 
 &  & 0.90 & 0.80 & 0.90 &  & 0.80 & 0.07 & 899.5 & 0.91 & 0.53 & 684.8 \\ \hline
\end{tabular}
\vspace{-0.2em}
\begin{tablenotes}
      \small
      \item  \hspace{30pt} * When the fraction of information $\omega$ reaches 0.5.
\end{tablenotes}
\end{table}

\section{Proof of Theorem 2}\label{Sup: proof-theorem2-boundary}
As discussed in main text, the inequality $\beta_L(\lambda_1, \lambda) \geq \alpha_L$ can be rewritten as 
 \begin{equation} \nonumber
 \Pr(\widehat{p}_{1,H} - \widehat{p}_{1,L} > \lambda_1 \mid \mathcal{S}_L) + \Pr(\widehat{p}_{1,H} - \widehat{p}_{1,L} \leq \lambda_1, \widehat{p}_{H} - \widehat{p}_{L} > \lambda|\mathcal{S}_L) \leq 1 - \alpha_L 
\end{equation}
Let $\sigma_L^2 = 2p_H(1-p_H)$. Define the standardized statistics as $Z_1 = \frac{\widehat{p}_{1,H} -\widehat{p}_{1,L}}{\sigma_L /\sqrt{n_1}}$ for the interim analysis and $Z = \frac{\widehat{p}_{H} -\widehat{p}_{L}}{\sigma_L /\sqrt{n}}$ for the final analysis under $\mathcal{S}_L$. The probability of incorrectly selecting $d_H$ under $\mathcal{S}_L$ can be expressed as:
 \begin{equation}  \nonumber
 \Pr\left(Z_1 > \frac{\lambda_1}{\sigma_L/\sqrt{n_1}}\mid \mathcal{S}_L\right) + \Pr\left(Z_1 \leq \frac{\lambda_1}{\sigma_L/\sqrt{n_1}}, Z > \frac{\lambda}{\sigma_L/\sqrt{n}}|\mathcal{S}_L\right) \leq 1 - \alpha_L,
\end{equation}
 Given that $Z_1 \sim N(0,1)$ and  $Z \sim N(0,1)$, the joint distribution can be expressed as
\begin{eqnarray} \nonumber
\begin{pmatrix}
       Z_1 \\
        Z
    \end{pmatrix}  \sim\ \mathcal{N}_2
\begin{pmatrix}
    \begin{pmatrix}
        0 \\
        0
    \end{pmatrix},
    \begin{pmatrix}
    1 & \ \ \ \ \rho   \\
    \rho & \ \ \ \ 1
    \end{pmatrix}
\end{pmatrix}.
\label{bivnormal}
\end{eqnarray}
Denote $\widehat{p}_{2,H}$ and $\widehat{p}_{2,L}$ as the estimated response rates from $(n-n_1)$ patients in each dose enrolled after interim analysis, where $\widehat{p}_{2,j} =\frac{1}{n-n_1} \sum_{i=n_1+1}^{n}Y_{i,j}$, $j={L,H}$. We then define an independent increment statistics $Z_2 = \frac{\widehat{p}_{2,H} -\widehat{p}_{2,L}}{\sigma_L /\sqrt{n -n_1}}$, which is independent of $Z_1$. The difference between response rates estimated on total sample size can be expressed as
$$\widehat{p}_{H} -\widehat{p}_{L} = \frac{n_1(\widehat{p}_{1,H} -\widehat{p}_{1,L}) + (n-n_1)(\widehat{p}_{2,H} -\widehat{p}_{2,L})}{n}.$$
Consequently, the statistic $Z$ in the final analysis can be written as 
\begin{eqnarray}  \nonumber
    \begin{aligned}
    Z  &= \frac{n_1(\widehat{p}_{1,H} -\widehat{p}_{1,L}) + (n-n_1)(\widehat{p}_{2,H} -\widehat{p}_{2,L})}{\sigma_L \sqrt{n}} \\
     & = \sqrt{\frac{n_1}{n}}Z_1 + \sqrt{\frac{n-n_1}{n}}Z_2.
    \end{aligned}
\end{eqnarray}
To compute the covariance between $Z_1$ and $Z$, we have:
\begin{eqnarray}  \nonumber
    \begin{aligned}
    Cov(Z_1, Z)  &= Cov(Z_1, \sqrt{\frac{n_1}{n}}Z_1 ) + Cov(Z_1, \sqrt{\frac{n-n_1}{n}}Z_2) \\
     & = \sqrt{\frac{n_1}{n}} Var(Z_1) \\
     & = \sqrt{\frac{n_1}{n}}.
    \end{aligned}
\end{eqnarray}
The correlation coefficient $\rho$ is given by
\begin{eqnarray}  \nonumber
    \begin{aligned}
    \rho  &= \frac{Cov(Z_1, Z)}{Var(Z_1)Var(Z)} \\
     & = \sqrt{\frac{n_1}{n}}.
    \end{aligned}
\end{eqnarray}
Thus, the joint distribution $[Z_1,Z]$ follows a bivariate normal distribution with correlation $\rho = \sqrt{\frac{n_1}{n}}=\sqrt{\omega}$.

According to O'Brien-Fleming error spending function \citep{O'BrienFleming1979, LanandDeMets1983}, the cumulative probability of incorrect selection spent at the interim analysis function is given by
$$\alpha^* = 2\Phi\left( \frac{\Phi^{-1}((1-\alpha_L)/2)}{\sqrt{\omega}}\right),$$
where $\Phi(\cdot)$ is the cumulative distribution (CDF) of the standard normal distribution, and $\Phi^{-1}(\cdot)$ is its inverse (the quantile function). 
The corresponding decision boundary $\lambda_1^*$ is calculated as: 
$$\lambda^*_1 = \Phi^{-1}(1-\alpha^*).$$

\section{\texorpdfstring{Approximation of PCS $\beta(\lambda_1,\lambda)$ under $\mathcal{S}_H$ in Two-Stage ROSE}{Approximation of PCS in Two-Stage ROSE}}
 \label{Sec: search_n_two_stage_ROSE}
In two-stage ROSE design, the PCS under scenario $\mathcal{S}_H$ is defined as 
\begin{eqnarray} \nonumber
    \beta_H(\lambda_1, \lambda) &=& \Pr(\text{Select } d_H \mid \mathcal{S}_H)\\ \nonumber
                            &=&  \Pr(\widehat{p}_{1,H} - \widehat{p}_{1,L} > \lambda_1 \mid \mathcal{S}_H) + \Pr(\widehat{p}_{1,H} - \widehat{p}_{1,L} \leq \lambda_1, \widehat{p}_{H} - \widehat{p}_{L} > \lambda|\mathcal{S}_H) 
\end{eqnarray}
Recall that the difference in estimated response rates between two doses can be approximated by a normal distribution. Let $\sigma_H^2 = (p_H - \delta)(1-p_H+\delta) +p_H(1-p_H)$, and define the standerdized statistics $Z_1^* = \frac{\widehat{p}_{1,H} - \widehat{p}_{1,L}-\delta}{\sigma_H/\sqrt{n_1}}$ and $Z^* = \frac{\widehat{p}_{H} - \widehat{p}_{L}-\delta}{\sigma_H/\sqrt{n}} $. Then $ \beta_H(\lambda_1, \lambda)$ can be approximated as 
\begin{eqnarray}  \label{eqn:approx_PCS_two_stage}
    \begin{aligned}
     1 - \Phi\left(\frac{\lambda^*_1\sigma_L/\sqrt{n_1}-\delta}{\sigma_H/\sqrt{n_1}} \mid \mathcal{S}_H\right) 
    + \Pr\left(Z_1^*  \leq \frac{\lambda^*_1\sigma_L/\sqrt{n_1}-\delta}{\sigma_H/\sqrt{n_1}}, Z^*  > \frac{\lambda^*\sigma_L/\sqrt{n}-\delta}{\sigma_H/\sqrt{n}} \mid  \mathcal{S}_H\right). 
    \end{aligned}
\end{eqnarray}
Here, $Z_1^*$ and $Z^*$ jointly follow a bivariate normal distribution with correlation $\rho=\sqrt{\omega}$, which can be proved using a process similar to that presented in \ref{Sup: proof-theorem2-boundary}. The inequality is evaluated under the assumption $\mathcal{S}_H$ with $p_L = p_H - \delta$, as it guarantees $\beta_H(\lambda_1, \lambda) \geq \alpha_H$  for all $p_L < p_H -\delta$.  

Given thresholds $\lambda^*_1$ and $\lambda^*$, we can numerically search for the smallest $n$ that satisfies $\beta_H(\lambda_1, \lambda) \geq \alpha_H$, where $\beta_H(\lambda_1, \lambda)$ is approximated by equation (\ref{eqn:approx_PCS_two_stage}).

\newpage

\section{ROSE design when randomization ratio is C}\label{sec:allocarion_ratio_C}
\subsection{One-stage ROSE}
Let $n_H$ and $n_L$ denote the sample size randomized to $d_H$ and $d_L$, respectively, and define the sample estimate of $p_j$ as $\widehat{p}_j = \frac{1}{n_j}\sum_{i=1}^{n_j} Y_{i,j}$, $j=L, H$. When the randomization ratio between $d_H$ and $d_L$ is $C:1$ (i.e., $n_H/n_L=C$), the sampling distribution of $\widehat{p}_H - \widehat{p}_L$ can be approximated by 
$$\widehat{p}_H - \widehat{p}_L \sim N(p_H-p_L,   \frac{p_L(1-p_L)}{n_L}+\frac{p_H(1-p_H)}{Cn_L}).$$
Given $\mathcal{S}_L$ and $\mathcal{S}_H$, the PCS under these scenarios of interest are respectively given by
\begin{align}
\beta_L(\lambda) &= \Pr(\text{Select } d_L \mid \mathcal{S}_L) \nonumber \\ 
                 &= \Pr(\widehat{p}_H - \widehat{p}_L \leq \lambda \mid \mathcal{S}_L) \nonumber \\
                 &\approx \Pr\left(\frac{ \sqrt{n_L}(\widehat{p}_H-\widehat{p}_L)}{\sqrt{ p_H(1-p_H)(1+1/C) }} \leq \frac{ \sqrt{n_L}\lambda}{\sqrt{ p_H(1-p_H)(1+1/C) }} \mid \mathcal{S}_L\right) \nonumber \\ 
                &\approx \Phi\left( \frac{ \sqrt{n_L}\lambda}{\sqrt{ p_H(1-p_H)(1+1/C) }} \mid \mathcal{S}_L \right),   \nonumber  \\
\intertext{and} 
\beta_H(\lambda) &= \Pr(\text{Select } d_H \mid \mathcal{S}_H) \nonumber \\ 
                 &= \Pr(\widehat{p}_H - \widehat{p}_L > \lambda \mid \mathcal{S}_H) \nonumber \\
                 &\approx 1 - \Phi\left( \frac{\sqrt{n_L}(\lambda - \delta)}{\sqrt{(p_H-\delta)(1-p_H+\delta) + \frac{p_H(1-p_H)}{C} }} \mid \mathcal{S}_H \right).  \nonumber
\end{align}
where $\Phi(\cdot)$ is the cumulative distribution function (CDF) of a $N(0,1)$ distribution. $\beta_H(\lambda)$ is derived under $\mathcal{S}_H$ with $p_L = p_H - \delta$. Given $p_H$ and $\delta$, ensuring adequate PCS under this setting guarantees sufficient PCS for all $p_L < p_H - \delta$. 

The ROSE design controls the PCS under $\mathcal{S}_L$ and $\mathcal{S}_H$ at prespecified levels of $\alpha_L$ and $\alpha_H$, respectively, by choosing an appropriate decision boundary $\lambda$, i.e., 
\begin{equation} \nonumber
\beta_L(\lambda) \geq \alpha_L ~\,\, \mathrm{and}\,\,~ \beta_H(\lambda) \geq \alpha_H.
\end{equation}
 Given a fixed $n$, this results in the following solution:
\begin{equation} \label{intervalsoluation_AR_C}  \nonumber
\lambda \in (\lambda_L \equiv \frac{\sqrt{p_H(1-p_H)(1+ \frac{1}{C}) }}{\sqrt{n_L}}\Phi^{-1}(\alpha_L), \, \lambda_H \equiv \frac{\sqrt{ (p_H-\delta)(1-p_H+\delta) + \frac{p_H(1-p_H)}{C} }}{\sqrt{n_L}}\Phi^{-1}(1-\alpha_H) + \delta)
\end{equation}
when $\lambda_L\le\lambda_H$. When $\lambda_L > \lambda_H$,  $\lambda$ has no solution, which means that the PCS under $\mathcal{S}_L$ and $\mathcal{S}_H$ cannot simultaneously reach their respective desired levels.

As proofed in section \ref{Sup: proof-theorem1-samplesize}, the sample size is minimized when \(\lambda_L = \lambda_H\), and is given by
   \begin{equation} \label{eqn:sample_size_AR_C}
        n_L = \left(\frac{\sqrt{p_H(1-p_H)(1+\frac{1}{C})}\Phi^{-1}(\alpha_L) - \sqrt{ (p_H - \delta)(1-p_H + \delta)+ \frac{p_H(1-p_H)}{C} }\Phi^{-1}(1-\alpha_H) }{\delta}\right)^2.
    \end{equation}
The decision boundary $\lambda$ corresponding to the minimal $n_L$ is given by
\begin{equation} \label{eqn:decision_boundary_AR_C}
\lambda = \frac{ \delta \sqrt{p_H(1-p_H)(1+\frac{1}{C})}\Phi^{-1}(\alpha_L)}{\sqrt{p_H(1-p_H)(1+\frac{1}{C})}\Phi^{-1}(\alpha_L) - \sqrt{(p_H-\delta)(1-p_H+\delta) + \frac{p_H(1-p_H)}{C} }\Phi^{-1}(1-\alpha_H)}.
\end{equation}

In practice, $n_L$ is rounded to its ceiling, $\lceil n_L \rceil$ and $n_H = \lceil Cn_L \rceil$.  

\subsection{Two-Stage ROSE}
Recall that $n_j$ represents the minimum total sample size for dose $d_j$, where $j=L,H$. Given the randomization ratio of $C:1$, we have $n_H = Cn_L$. Let $n_{1,j}$ denote the number of patients enrolled in $d_j$ at the interim monitoring point. The corresponding sample size for dose $d_j$ at interim is $n_{1,j} = \omega n_{j}$, where $\omega$ ds the pre-specified information fraction (i.e., the proportion of the total planned sample size at which the interim analysis is conducted).  The estimated response rate for dose $j$ at the interim analysis is denoted by $\widehat{p}_{1,j}$, where $\widehat{p}_{1,j} =\frac{1}{n_{1,j}} \sum_{i=1}^{n_{1,j}}Y_{i,j}$, $j={L,H}$. Recall that $\widehat{p}_j$ represents the estimated response rate for dose $j$ based on total sample size $n_j$ for dose $d_j$. 

Given the interim information fraction $\omega$, our objective is to determine the smallest sample size $n_j$ and decision boundaries, $\lambda_1$ and $\lambda$, that ensure the probabilities of correct selection under both scenarios of interest achieve pre-specified desired levels, i.e.,  $\beta_L(\lambda_1, \lambda) \geq \alpha_L$ and $\beta_H(\lambda_1, \lambda) \geq \alpha_H$. The calculation procedure follows the same rationale as in the 1:1 allocation case described in the main text, with adjustments for the unequal allocation. Specifically, $\sigma_L^2$ is defined as $\sigma_L^2 = p_H(1-p_H)(1+\frac{1}{C})$, and the standerdized decision boundaries are defined as $\lambda^*_1 = \frac{\lambda_1}{\sigma_L/\sqrt{n_{1,L}}}$ and $\lambda^* = \frac{\lambda}{\sigma_L/\sqrt{n_L}}.$ The interim standardized boundary $\lambda^*_1$ remains the same as in the  equal allocation case and is given by 
$$\lambda^*_1 =\Phi^{-1}\left\{1- 2\Phi\left( \frac{\Phi^{-1}((1-\alpha_L)/2)}{\sqrt{\omega}}\right) \right\},$$

Given the thresholds $\lambda^*_1$ and $\lambda^*$, we can numerically search for the smallest $n_L$ that satisfies $\beta_H(\lambda_1, \lambda) \geq \alpha_H$. Once the total sample size for dose $d_L$ is determined, the first-stage sample size for dose $d_L$, $n_{1,L}$, is obtained by $\lceil \omega n_L \rceil$, the nearest integer that $\geq \omega n_L $. The corresponding sample sizes at interim and final analyses for dose $d_H$ are $n_{1,H} = \lceil Cn_{1,L} \rceil$ and $n_H = \lceil Cn_L \rceil$, respectively. The decision boundaries $\lambda_1$ and $\lambda$, can be calculated as  $\lambda_1 = \lambda^*_1\sigma_L/\sqrt{n_{1,L}}$ and  $\lambda = \lambda^*\sigma_L/\sqrt{n_L}$. 

\begin{table}[H]
\centering
\caption{Decision boundaries and sample sizes per dose arm for ROSE designs when $\delta=0.1$ and $C=2$.} 
\label{tab:Bounderies_sample_size_delta_0.1_C_2}
\setlength{\tabcolsep}{8pt}
\begin{tabular}{ccccccp{0.05cm}ccccccc}
\hline
\multirow{2}{*}{$p_H$} & \multirow{2}{*}{$\alpha_L$} & \multirow{2}{*}{$\alpha_H$} & \multicolumn{3}{c}{No interim} &  & \multicolumn{6}{c}{With one interim$^*$} \\ \cline{4-6} \cline{8-13} 
 &  &  & $\lambda$ & $n_L$ & $n_H$ &  & $\lambda_1$ & $n_{1,L}$ & $n_{1,H}$ & $\lambda$ & $n_L$ & $n_H$ \\ \hline
\multirow{10}{*}{0.3} & \multirow{2}{*}{0.60} & 0.60 & 0.052 & 8 & 15 &  & 0.182 & 5 & 10 & 0.074 & 10 & 20 \\ \cline{3-13} 
 &  & 0.70 & 0.035 & 17 & 34 &  & 0.129 & 10 & 20 & 0.052 & 20 & 40 \\ \cline{2-13} 
 & \multirow{2}{*}{0.65} & 0.65 & 0.052 & 18 & 35 &  & 0.158 & 10 & 20 & 0.066 & 20 & 40 \\ \cline{3-13} 
 &  & 0.75 & 0.038 & 32 & 64 &  & 0.118 & 18 & 36 & 0.049 & 36 & 72 \\ \cline{2-13} 
 & \multirow{2}{*}{0.70} & 0.70 & 0.052 & 32 & 64 &  & 0.141 & 18 & 36 & 0.060 & 36 & 72 \\ \cline{3-13} 
 &  & 0.80 & 0.040 & 53 & 106 &  & 0.111 & 29 & 58 & 0.047 & 58 & 116 \\ \cline{2-13} 
 & \multirow{2}{*}{0.75} & 0.75 & 0.052 & 53 & 106 &  & 0.131 & 29 & 57 & 0.057 & 57 & 114 \\ \cline{3-13} 
 &  & 0.85 & 0.042 & 84 & 167 &  & 0.105 & 45 & 89 & 0.046 & 89 & 178 \\ \cline{2-13} 
 & \multirow{2}{*}{0.80} & 0.80 & 0.052 & 83 & 165 &  & 0.125 & 44 & 87 & 0.055 & 87 & 174 \\ \cline{3-13} 
 &  & 0.90 & 0.042 & 129 & 257 &  & 0.100 & 68 & 135 & 0.044 & 135 & 270 \\ \hline
\multirow{10}{*}{0.4} & \multirow{2}{*}{0.60} & 0.60 & 0.051 & 9 & 18 &  & 0.178 & 6 & 11 & 0.075 & 11 & 22 \\ \cline{3-13} 
 &  & 0.70 & 0.034 & 21 & 42 &  & 0.126 & 12 & 24 & 0.051 & 24 & 48 \\ \cline{2-13} 
 & \multirow{2}{*}{0.65} & 0.65 & 0.051 & 21 & 41 &  & 0.154 & 12 & 23 & 0.065 & 23 & 46 \\ \cline{3-13} 
 &  & 0.75 & 0.037 & 39 & 77 &  & 0.114 & 22 & 43 & 0.048 & 43 & 86 \\ \cline{2-13} 
 & \multirow{2}{*}{0.70} & 0.70 & 0.051 & 38 & 76 &  & 0.140 & 21 & 42 & 0.059 & 42 & 84 \\ \cline{3-13} 
 &  & 0.80 & 0.039 & 64 & 128 &  & 0.108 & 35 & 69 & 0.046 & 69 & 138 \\ \cline{2-13} 
 & \multirow{2}{*}{0.75} & 0.75 & 0.051 & 63 & 126 &  & 0.130 & 34 & 67 & 0.056 & 67 & 134 \\ \cline{3-13} 
 &  & 0.85 & 0.040 & 101 & 201 &  & 0.104 & 53 & 106 & 0.045 & 106 & 212 \\ \cline{2-13} 
 & \multirow{2}{*}{0.80} & 0.80 & 0.051 & 98 & 196 &  & 0.123 & 52 & 103 & 0.054 & 103 & 206 \\ \cline{3-13} 
 &  & 0.90 & 0.041 & 155 & 309 &  & 0.098 & 81 & 161 & 0.043 & 161 & 322 \\ \hline
\multirow{10}{*}{0.5} & \multirow{2}{*}{0.60} & 0.60 & 0.050 & 10 & 19 &  & 0.181 & 6 & 11 & 0.076 & 11 & 22 \\ \cline{3-13} 
 &  & 0.70 & 0.033 & 23 & 45 &  & 0.123 & 13 & 25 & 0.051 & 25 & 50 \\ \cline{2-13} 
 & \multirow{2}{*}{0.65} & 0.65 & 0.050 & 22 & 44 &  & 0.151 & 13 & 25 & 0.064 & 25 & 50 \\ \cline{3-13} 
 &  & 0.75 & 0.037 & 42 & 83 &  & 0.114 & 23 & 46 & 0.047 & 46 & 92 \\ \cline{2-13} 
 & \multirow{2}{*}{0.70} & 0.70 & 0.050 & 41 & 82 &  & 0.139 & 22 & 44 & 0.059 & 44 & 88 \\ \cline{3-13} 
 &  & 0.80 & 0.039 & 69 & 138 &  & 0.108 & 37 & 74 & 0.045 & 74 & 148 \\ \cline{2-13} 
 & \multirow{2}{*}{0.75} & 0.75 & 0.050 & 68 & 135 &  & 0.129 & 36 & 72 & 0.055 & 72 & 144 \\ \cline{3-13} 
 &  & 0.85 & 0.040 & 108 & 216 &  & 0.102 & 57 & 114 & 0.044 & 114 & 228 \\ \cline{2-13} 
 & \multirow{2}{*}{0.80} & 0.80 & 0.050 & 105 & 210 &  & 0.122 & 55 & 110 & 0.053 & 110 & 220 \\ \cline{3-13} 
 &  & 0.90 & 0.040 & 167 & 333 &  & 0.097 & 87 & 173 & 0.043 & 173 & 346 \\ \hline
\end{tabular}
\vspace{-0.2em}
\begin{tablenotes}
      \small
      \item  * When the fraction of information $\omega$ reaches 0.5.
\end{tablenotes}
\end{table}

\begin{table}[H]
\centering
\caption{Decision boundaries and sample sizes per dose arm for ROSE designs when $\delta=0.15$ and $C=2$.} 
\label{tab:Bounderies_sample_size_delta_0.15_C_2}
\setlength{\tabcolsep}{8pt}
\begin{tabular}{ccccccp{0.05cm}ccccccc}
\hline
\multirow{2}{*}{$p_H$} & \multirow{2}{*}{$\alpha_L$} & \multirow{2}{*}{$\alpha_H$} & \multicolumn{3}{c}{No interim} &  & \multicolumn{6}{c}{With one interim$^*$} \\ \cline{4-6} \cline{8-13} 
 &  &  & $\lambda$ & $n_L$ & $n_H$ &  & $\lambda_1$ & $n_{1,L}$ & $n_{1,H}$ & $\lambda$ & $n_L$ & $n_H$ \\ \hline
\multirow{10}{*}{0.3} & \multirow{2}{*}{0.60} & 0.60 & 0.081 & 4 & 7 &  & 0.235 & 3 & 5 & 0.104 & 5 & 10 \\ \cline{3-13} 
 &  & 0.70 & 0.054 & 7 & 14 &  & 0.182 & 5 & 9 & 0.077 & 9 & 18 \\ \cline{2-13} 
 & \multirow{2}{*}{0.65} & 0.65 & 0.081 & 8 & 15 &  & 0.224 & 5 & 9 & 0.098 & 9 & 18 \\ \cline{3-13} 
 &  & 0.75 & 0.060 & 14 & 27 &  & 0.177 & 8 & 16 & 0.073 & 16 & 32 \\ \cline{2-13} 
 & \multirow{2}{*}{0.70} & 0.70 & 0.081 & 14 & 27 &  & 0.212 & 8 & 16 & 0.090 & 16 & 32 \\ \cline{3-13} 
 &  & 0.80 & 0.063 & 22 & 44 &  & 0.166 & 13 & 25 & 0.072 & 25 & 50 \\ \cline{2-13} 
 & \multirow{2}{*}{0.75} & 0.75 & 0.081 & 23 & 45 &  & 0.196 & 13 & 25 & 0.086 & 25 & 50 \\ \cline{3-13} 
 &  & 0.85 & 0.065 & 35 & 69 &  & 0.162 & 19 & 37 & 0.071 & 37 & 74 \\ \cline{2-13} 
 & \multirow{2}{*}{0.80} & 0.80 & 0.081 & 35 & 69 &  & 0.190 & 19 & 37 & 0.084 & 37 & 74 \\ \cline{3-13} 
 &  & 0.90 & 0.065 & 53 & 106 &  & 0.157 & 28 & 56 & 0.069 & 56 & 112 \\ \hline
\multirow{10}{*}{0.4} & \multirow{2}{*}{0.60} & 0.60 & 0.078 & 4 & 8 &  & 0.251 & 3 & 5 & 0.111 & 5 & 10 \\ \cline{3-13} 
 &  & 0.70 & 0.051 & 9 & 18 &  & 0.178 & 6 & 11 & 0.075 & 11 & 22 \\ \cline{2-13} 
 & \multirow{2}{*}{0.65} & 0.65 & 0.078 & 9 & 18 &  & 0.218 & 6 & 11 & 0.094 & 11 & 22 \\ \cline{3-13} 
 &  & 0.75 & 0.057 & 17 & 33 &  & 0.169 & 10 & 19 & 0.072 & 19 & 38 \\ \cline{2-13} 
 & \multirow{2}{*}{0.70} & 0.70 & 0.078 & 17 & 33 &  & 0.214 & 9 & 18 & 0.090 & 18 & 36 \\ \cline{3-13} 
 &  & 0.80 & 0.060 & 28 & 55 &  & 0.165 & 15 & 30 & 0.070 & 30 & 60 \\ \cline{2-13} 
 & \multirow{2}{*}{0.75} & 0.75 & 0.078 & 27 & 54 &  & 0.195 & 15 & 29 & 0.085 & 29 & 58 \\ \cline{3-13} 
 &  & 0.85 & 0.062 & 43 & 86 &  & 0.158 & 23 & 46 & 0.068 & 46 & 92 \\ \cline{2-13} 
 & \multirow{2}{*}{0.80} & 0.80 & 0.078 & 42 & 84 &  & 0.185 & 23 & 45 & 0.082 & 45 & 90 \\ \cline{3-13} 
 &  & 0.90 & 0.062 & 66 & 132 &  & 0.150 & 35 & 69 & 0.066 & 69 & 138 \\ \hline
\multirow{10}{*}{0.5} & \multirow{2}{*}{0.60} & 0.60 & 0.076 & 5 & 9 &  & 0.257 & 3 & 5 & 0.113 & 5 & 10 \\ \cline{3-13} 
 &  & 0.70 & 0.050 & 10 & 20 &  & 0.181 & 6 & 11 & 0.076 & 11 & 22 \\ \cline{2-13} 
 & \multirow{2}{*}{0.65} & 0.65 & 0.076 & 10 & 20 &  & 0.223 & 6 & 11 & 0.096 & 11 & 22 \\ \cline{3-13} 
 &  & 0.75 & 0.056 & 19 & 37 &  & 0.173 & 10 & 20 & 0.071 & 20 & 40 \\ \cline{2-13} 
 & \multirow{2}{*}{0.70} & 0.70 & 0.076 & 18 & 36 &  & 0.207 & 10 & 20 & 0.087 & 20 & 40 \\ \cline{3-13} 
 &  & 0.80 & 0.059 & 30 & 60 &  & 0.159 & 17 & 33 & 0.068 & 33 & 66 \\ \cline{2-13} 
 & \multirow{2}{*}{0.75} & 0.75 & 0.076 & 30 & 59 &  & 0.193 & 16 & 32 & 0.083 & 32 & 64 \\ \cline{3-13} 
 &  & 0.85 & 0.060 & 48 & 95 &  & 0.154 & 25 & 50 & 0.066 & 50 & 100 \\ \cline{2-13} 
 & \multirow{2}{*}{0.80} & 0.80 & 0.076 & 46 & 92 &  & 0.185 & 24 & 48 & 0.081 & 48 & 96 \\ \cline{3-13} 
 &  & 0.90 & 0.061 & 73 & 145 &  & 0.147 & 38 & 76 & 0.064 & 76 & 152 \\ \hline
\end{tabular}
\vspace{-0.2em}
\begin{tablenotes}
      \small
      \item  * When the fraction of information $\omega$ reaches 0.5.
\end{tablenotes}
\end{table}

\newpage
\begin{table}[H]
\centering
\caption{Simulation results of one-stage and two-stage ROSE design when $\delta =0.1$ and $C=2$.} 
\label{tab:sim_result_design_delta_0.1_C_2}
\begin{tabular}{cccccccccccc}
\hline
\multirow{3}{*}{$p_H$} & \multirow{3}{*}{$\alpha_L$} & \multirow{3}{*}{$\alpha_H$} & \multicolumn{2}{c}{No interim} &  & \multicolumn{6}{c}{With one interim$^*$} \\ \cline{4-5} \cline{7-12} 
 &  &   & \multicolumn{2}{c}{PCS} &  & \multicolumn{3}{c}{$p_L=0.3$} & \multicolumn{3}{c}{$p_L=0.2$} \\ \cline{4-5} \cline{7-12} 
 &   &  & $p_L=0.3$ & $p_L=0.2$ &  &PCS  &PET  &EN$^{\dagger}$   &PCS  &PET  &EN$^{\dagger}$  \\ \hline
\multirow{10}{*}{0.3} & \multirow{2}{*}{0.60} & 0.60 & 0.59 & 0.62 &  & 0.58 & 0.29 & 25.7 & 0.63 & 0.43 & 23.6 \\ \cline{3-12} 
 &  & 0.70 & 0.62 & 0.67 &  & 0.61 & 0.25 & 52.6 & 0.70 & 0.46 & 46.1 \\ \cline{2-12} 
 & \multirow{2}{*}{0.65} & 0.65 & 0.64 & 0.67 &  & 0.65 & 0.16 & 55.1 & 0.66 & 0.33 & 50.0 \\ \cline{3-12} 
 &  & 0.75 & 0.65 & 0.75 &  & 0.66 & 0.17 & 98.7 & 0.76 & 0.43 & 84.9 \\ \cline{2-12} 
 & \multirow{2}{*}{0.70} & 0.70 & 0.72 & 0.69 &  & 0.72 & 0.13 & 101.2 & 0.69 & 0.33 & 90.2 \\ \cline{3-12} 
 &  & 0.80 & 0.71 & 0.80 &  & 0.70 & 0.15 & 161.1 & 0.79 & 0.45 & 134.8 \\ \cline{2-12} 
 & \multirow{2}{*}{0.75} & 0.75 & 0.75 & 0.76 &  & 0.75 & 0.09 & 163.0 & 0.75 & 0.37 & 139.9 \\ \cline{3-12} 
 &  & 0.85 & 0.74 & 0.86 &  & 0.76 & 0.09 & 254.5 & 0.84 & 0.46 & 205.9 \\ \cline{2-12} 
 & \multirow{2}{*}{0.80} & 0.80 & 0.81 & 0.80 &  & 0.80 & 0.07 & 252.0 & 0.81 & 0.39 & 210.2 \\ \cline{3-12} 
 &  & 0.90 & 0.80 & 0.90 &  & 0.79 & 0.06 & 391.9 & 0.91 & 0.50 & 304.6 \\ \hline
 &  &   & \multicolumn{2}{c}{PCS} &  & \multicolumn{3}{c}{$p_L=0.4$} & \multicolumn{3}{c}{$p_L=0.3$} \\ \cline{4-5} \cline{7-12} 
 &   &  & $p_L=0.4$ & $p_L=0.3$ &  &PCS  &PET  &EN$^{\dagger}$   &PCS  &PET  &EN$^{\dagger}$  \\ \hline 
\multirow{10}{*}{0.4} & \multirow{2}{*}{0.60} & 0.60 & 0.54 & 0.65 &  & 0.57 & 0.26 & 28.8 & 0.64 & 0.42 & 26.3 \\ \cline{3-12} 
 &  & 0.70 & 0.61 & 0.70 &  & 0.61 & 0.21 & 64.6 & 0.70 & 0.40 & 57.6 \\ \cline{2-12} 
 & \multirow{2}{*}{0.65} & 0.65 & 0.64 & 0.67 &  & 0.67 & 0.18 & 62.8 & 0.63 & 0.36 & 56.7 \\ \cline{3-12} 
 &  & 0.75 & 0.65 & 0.76 &  & 0.66 & 0.19 & 116.6 & 0.74 & 0.47 & 99.1 \\ \cline{2-12} 
 & \multirow{2}{*}{0.70} & 0.70 & 0.68 & 0.72 &  & 0.68 & 0.16 & 116.0 & 0.72 & 0.40 & 100.7 \\ \cline{3-12} 
 &  & 0.80 & 0.72 & 0.79 &  & 0.70 & 0.14 & 193.0 & 0.80 & 0.45 & 160.8 \\ \cline{2-12} 
 & \multirow{2}{*}{0.75} & 0.75 & 0.74 & 0.75 &  & 0.75 & 0.10 & 191.3 & 0.75 & 0.37 & 163.6 \\ \cline{3-12} 
 &  & 0.85 & 0.74 & 0.86 &  & 0.75 & 0.10 & 302.4 & 0.85 & 0.46 & 244.9 \\ \cline{2-12} 
 & \multirow{2}{*}{0.80} & 0.80 & 0.81 & 0.80 &  & 0.80 & 0.07 & 298.8 & 0.79 & 0.38 & 250.4 \\ \cline{3-12} 
 &  & 0.90 & 0.81 & 0.90 &  & 0.79 & 0.07 & 466.7 & 0.90 & 0.52 & 358.0 \\ \hline
 &  &   & \multicolumn{2}{c}{PCS} &  & \multicolumn{3}{c}{$p_L=0.5$} & \multicolumn{3}{c}{$p_L=0.4$} \\ \cline{4-5} \cline{7-12} 
 &   &  & $p_L=0.5$ & $p_L=0.4$ &  &PCS  &PET  &EN$^{\dagger}$   &PCS  &PET  &EN$^{\dagger}$  \\ \hline
\multirow{10}{*}{0.5} & \multirow{2}{*}{0.60} & 0.60 & 0.60 & 0.61 &  & 0.58 & 0.25 & 28.9 & 0.63 & 0.40 & 26.5 \\ \cline{3-12} 
 &  & 0.70 & 0.62 & 0.68 &  & 0.59 & 0.25 & 65.8 & 0.71 & 0.47 & 57.7 \\ \cline{2-12} 
 & \multirow{2}{*}{0.65} & 0.65 & 0.67 & 0.63 &  & 0.67 & 0.18 & 68.5 & 0.65 & 0.39 & 60.6 \\ \cline{3-12} 
 &  & 0.75 & 0.65 & 0.76 &  & 0.66 & 0.18 & 125.9 & 0.75 & 0.45 & 107.3 \\ \cline{2-12} 
 & \multirow{2}{*}{0.70} & 0.70 & 0.71 & 0.68 &  & 0.71 & 0.13 & 123.5 & 0.68 & 0.36 & 108.5 \\ \cline{3-12} 
 &  & 0.80 & 0.71 & 0.80 &  & 0.68 & 0.16 & 204.2 & 0.81 & 0.49 & 167.2 \\ \cline{2-12} 
 & \multirow{2}{*}{0.75} & 0.75 & 0.75 & 0.74 &  & 0.75 & 0.10 & 205.7 & 0.76 & 0.38 & 174.9 \\ \cline{3-12} 
 &  & 0.85 & 0.75 & 0.86 &  & 0.76 & 0.11 & 323.8 & 0.84 & 0.49 & 258.6 \\ \cline{2-12} 
 & \multirow{2}{*}{0.80} & 0.80 & 0.80 & 0.80 &  & 0.80 & 0.07 & 318.5 & 0.82 & 0.39 & 265.2 \\ \cline{3-12} 
 &  & 0.90 & 0.80 & 0.91 &  & 0.80 & 0.07 & 501.9 & 0.91 & 0.52 & 384.4 \\ \hline
\end{tabular}
\vspace{-0.2em}
\begin{tablenotes}
      \small
      \item  \hspace{18pt} * When the fraction of information $\omega$ reaches 0.5.
      \item \hspace{18pt} ${\dagger}$ Here, EN refers to the total expected sample size, not per dose arm. 
\end{tablenotes}
\end{table}

\begin{table}[H]
\centering
\caption{Simulation results of one-stage and two-stage ROSE design when $\delta =0.15$ and $C=2$.}
\label{tab:sim_result_design_delta_0.15_C_2}
\begin{tabular}{cccccccccccc}
\hline
\multirow{3}{*}{$p_H$} & \multirow{3}{*}{$\alpha_L$} & \multirow{3}{*}{$\alpha_H$} & \multicolumn{2}{c}{No interim} &  & \multicolumn{6}{c}{With one interim$^*$} \\ \cline{4-5} \cline{7-12} 
 &  &   & \multicolumn{2}{c}{PCS} &  & \multicolumn{3}{c}{$p_L=0.3$} & \multicolumn{3}{c}{$p_L=0.15$} \\ \cline{4-5} \cline{7-12} 
 &   &  & $p_L=0.3$ & $p_L=0.15$ &  &PCS  &PET  &EN$^{\dagger}$   &PCS  &PET  &EN$^{\dagger}$  \\ \hline
\multirow{10}{*}{0.3} & \multirow{2}{*}{0.60} & 0.60 & 0.63 & 0.61 &  & 0.63 & 0.24 & 13.3 & 0.57 & 0.34 & 12.6 \\ \cline{3-12} 
 &  & 0.70 & 0.55 & 0.74 &  & 0.60 & 0.24 & 23.8 & 0.71 & 0.47 & 21.0 \\ \cline{2-12} 
 & \multirow{2}{*}{0.65} & 0.65 & 0.63 & 0.66 &  & 0.62 & 0.20 & 24.5 & 0.70 & 0.35 & 22.4 \\ \cline{3-12} 
 &  & 0.75 & 0.67 & 0.75 &  & 0.64 & 0.22 & 42.7 & 0.77 & 0.50 & 35.9 \\ \cline{2-12} 
 & \multirow{2}{*}{0.70} & 0.70 & 0.68 & 0.71 &  & 0.68 & 0.13 & 44.8 & 0.75 & 0.35 & 39.6 \\ \cline{3-12} 
 &  & 0.80 & 0.68 & 0.82 &  & 0.70 & 0.14 & 69.9 & 0.81 & 0.43 & 59.1 \\ \cline{2-12} 
 & \multirow{2}{*}{0.75} & 0.75 & 0.75 & 0.76 &  & 0.75 & 0.11 & 70.8 & 0.76 & 0.41 & 60.0 \\ \cline{3-12} 
 &  & 0.85 & 0.76 & 0.85 &  & 0.76 & 0.12 & 104.5 & 0.84 & 0.50 & 83.7 \\ \cline{2-12} 
 & \multirow{2}{*}{0.80} & 0.80 & 0.81 & 0.80 &  & 0.80 & 0.08 & 106.5 & 0.81 & 0.41 & 88.2 \\ \cline{3-12} 
 &  & 0.90 & 0.78 & 0.91 &  & 0.79 & 0.08 & 161.6 & 0.91 & 0.50 & 126.0 \\ \hline
 &  &   & \multicolumn{2}{c}{PCS} &  & \multicolumn{3}{c}{$p_L=0.4$} & \multicolumn{3}{c}{$p_L=0.25$} \\ \cline{4-5} \cline{7-12} 
 &   &  & $p_L=0.4$ & $p_L=0.25$ &  &PCS  &PET  &EN$^{\dagger}$   &PCS  &PET  &EN$^{\dagger}$  \\ \hline
\multirow{10}{*}{0.4} & \multirow{2}{*}{0.60} & 0.60 & 0.58 & 0.63 &  & 0.61 & 0.29 & 13.0 & 0.59 & 0.42 & 12.1 \\ \cline{3-12} 
 &  & 0.70 & 0.56 & 0.75 &  & 0.58 & 0.26 & 28.9 & 0.74 & 0.50 & 24.9 \\ \cline{2-12} 
 & \multirow{2}{*}{0.65} & 0.65 & 0.66 & 0.65 &  & 0.69 & 0.17 & 30.3 & 0.63 & 0.36 & 27.2 \\ \cline{3-12} 
 &  & 0.75 & 0.64 & 0.77 &  & 0.63 & 0.20 & 51.4 & 0.77 & 0.45 & 44.4 \\ \cline{2-12} 
 & \multirow{2}{*}{0.70} & 0.70 & 0.71 & 0.70 &  & 0.70 & 0.17 & 49.3 & 0.70 & 0.42 & 42.6 \\ \cline{3-12} 
 &  & 0.80 & 0.69 & 0.79 &  & 0.70 & 0.17 & 82.6 & 0.81 & 0.51 & 66.9 \\ \cline{2-12} 
 & \multirow{2}{*}{0.75} & 0.75 & 0.76 & 0.73 &  & 0.74 & 0.10 & 82.6 & 0.78 & 0.38 & 70.7 \\ \cline{3-12} 
 &  & 0.85 & 0.75 & 0.85 &  & 0.76 & 0.10 & 131.4 & 0.84 & 0.46 & 106.3 \\ \cline{2-12} 
 & \multirow{2}{*}{0.80} & 0.80 & 0.80 & 0.81 &  & 0.81 & 0.06 & 130.7 & 0.81 & 0.37 & 110.2 \\ \cline{3-12} 
 &  & 0.90 & 0.81 & 0.90 &  & 0.82 & 0.06 & 200.8 & 0.90 & 0.47 & 158.1 \\ \hline
 &  &   & \multicolumn{2}{c}{PCS} &  & \multicolumn{3}{c}{$p_L=0.5$} & \multicolumn{3}{c}{$p_L=0.35$} \\ \cline{4-5} \cline{7-12} 
 &   &  & $p_L=0.5$ & $p_L=0.35$ &  &PCS  &PET  &EN$^{\dagger}$   &PCS  &PET  &EN$^{\dagger}$  \\ \hline
\multirow{10}{*}{0.5} & \multirow{2}{*}{0.60} & 0.60 & 0.64 & 0.58 &  & 0.61 & 0.30 & 12.9 & 0.61 & 0.46 & 11.8 \\ \cline{3-12} 
 &  & 0.70 & 0.55 & 0.75 &  & 0.59 & 0.26 & 28.9 & 0.72 & 0.49 & 25.2 \\ \cline{2-12} 
 & \multirow{2}{*}{0.65} & 0.65 & 0.65 & 0.66 &  & 0.67 & 0.19 & 30.0 & 0.62 & 0.36 & 27.2 \\ \cline{3-12} 
 &  & 0.75 & 0.65 & 0.76 &  & 0.63 & 0.19 & 54.4 & 0.76 & 0.45 & 46.6 \\ \cline{2-12} 
 & \multirow{2}{*}{0.70} & 0.70 & 0.68 & 0.72 &  & 0.70 & 0.13 & 56.2 & 0.70 & 0.36 & 49.3 \\ \cline{3-12} 
 &  & 0.80 & 0.70 & 0.80 &  & 0.68 & 0.16 & 91.2 & 0.81 & 0.50 & 74.3 \\ \cline{2-12} 
 & \multirow{2}{*}{0.75} & 0.75 & 0.76 & 0.73 &  & 0.76 & 0.10 & 91.4 & 0.74 & 0.37 & 78.2 \\ \cline{3-12} 
 &  & 0.85 & 0.76 & 0.84 &  & 0.73 & 0.12 & 141.3 & 0.86 & 0.51 & 111.4 \\ \cline{2-12} 
 & \multirow{2}{*}{0.80} & 0.80 & 0.82 & 0.78 &  & 0.79 & 0.08 & 138.5 & 0.81 & 0.42 & 113.8 \\ \cline{3-12} 
 &  & 0.90 & 0.81 & 0.90 &  & 0.79 & 0.07 & 220.5 & 0.91 & 0.50 & 171.3 \\ \hline
\end{tabular}
\vspace{-0.2em}
\begin{tablenotes}
      \small
      \item  \hspace{15pt} * When the fraction of information $\omega$ reaches 0.5.
      \item \hspace{15pt} ${\dagger}$ Here, EN refers to the total expected sample size, not per dose arm. 
\end{tablenotes}
\end{table}

\newpage
\section{Decision boundary, sample size per
arm, and operating characteristics of eROSE design based on binomial distribution} \label{sec:eROSE_boundary}

\begin{table}[H]
\centering
\caption{Decision boundaries and sample sizes per dose arm for eROSE designs when $\delta=0.1$.} 
\label{tab:Bounderies_sample_size_eROSE_delta_0.1}
\setlength{\tabcolsep}{10pt}
\begin{tabular}{ccccc@{}ccccc}
\hline
\multirow{2}{*}{$p_H$} & \multirow{2}{*}{$\alpha_L$} & \multirow{2}{*}{$\alpha_H$} & \multicolumn{2}{c}{No interim} &  & \multicolumn{4}{c}{With one interim$^*$} \\ \cline{4-5} \cline{7-10} 
 &  &  & $\lambda$ & $n$ &  & $\lambda_1$ & $n_1$ & $\lambda$ & $n$ \\ \hline
\multirow{10}{*}{0.3} & \multirow{2}{*}{0.60} & 0.60 & 0.044 & 23 &  & 0.1 & 10 & 0.054 & 19 \\ \cline{3-10} 
 &  & 0.70 & 0.03 & 34 &  & 0.126 & 16 & 0.034 & 31 \\ \cline{2-10} 
 & \multirow{2}{*}{0.65} & 0.65 & 0.036 & 28 &  & 0.112 & 18 & 0.058 & 35 \\ \cline{3-10} 
 &  & 0.75 & 0.036 & 56 &  & 0.112 & 27 & 0.038 & 53 \\ \cline{2-10} 
 & \multirow{2}{*}{0.70} & 0.70 & 0.044 & 47 &  & 0.138 & 29 & 0.052 & 58 \\ \cline{3-10} 
 &  & 0.80 & 0.038 & 82 &  & 0.118 & 43 & 0.036 & 85 \\ \cline{2-10} 
 & \multirow{2}{*}{0.75} & 0.75 & 0.05 & 83 &  & 0.126 & 40 & 0.05 & 80 \\ \cline{3-10} 
 &  & 0.85 & 0.04 & 126 &  & 0.1 & 61 & 0.042 & 122 \\ \cline{2-10} 
 & \multirow{2}{*}{0.80} & 0.80 & 0.05 & 122 &  & 0.118 & 60 & 0.052 & 119 \\ \cline{3-10} 
 &  & 0.90 & 0.04 & 180 &  & 0.096 & 95 & 0.044 & 190 \\ \hline
\multirow{10}{*}{0.4} & \multirow{2}{*}{0.60} & 0.60 & 0.042 & 24 &  & 0.182 & 11 & 0.046 & 22 \\ \cline{3-10} 
 &  & 0.70 & 0.028 & 37 &  & 0.118 & 17 & 0.032 & 33 \\ \cline{2-10} 
 & \multirow{2}{*}{0.65} & 0.65 & 0.036 & 29 &  & 0.15 & 20 & 0.052 & 39 \\ \cline{3-10} 
 &  & 0.75 & 0.034 & 60 &  & 0.122 & 33 & 0.032 & 65 \\ \cline{2-10} 
 & \multirow{2}{*}{0.70} & 0.70 & 0.048 & 63 &  & 0.134 & 30 & 0.05 & 60 \\ \cline{3-10} 
 &  & 0.80 & 0.036 & 88 &  & 0.104 & 49 & 0.042 & 97 \\ \cline{2-10} 
 & \multirow{2}{*}{0.75} & 0.75 & 0.046 & 88 &  & 0.124 & 49 & 0.052 & 97 \\ \cline{3-10} 
 &  & 0.85 & 0.038 & 137 &  & 0.096 & 73 & 0.042 & 145 \\ \cline{2-10} 
 & \multirow{2}{*}{0.80} & 0.80 & 0.05 & 143 &  & 0.13 & 70 & 0.05 & 140 \\ \cline{3-10} 
 &  & 0.90 & 0.04 & 210 &  & 0.1 & 111 & 0.042 & 221 \\ \hline
\multirow{10}{*}{0.5} & \multirow{2}{*}{0.60} & 0.60 & 0.042 & 24 &  & 0.182 & 11 & 0.046 & 22 \\ \cline{3-10} 
 &  & 0.70 & 0.028 & 38 &  & 0.118 & 17 & 0.032 & 33 \\ \cline{2-10} 
 & \multirow{2}{*}{0.65} & 0.65 & 0.034 & 30 &  & 0.15 & 20 & 0.05 & 40 \\ \cline{3-10} 
 &  & 0.75 & 0.032 & 63 &  & 0.13 & 31 & 0.034 & 61 \\ \cline{2-10} 
 & \multirow{2}{*}{0.70} & 0.70 & 0.048 & 65 &  & 0.13 & 31 & 0.05 & 61 \\ \cline{3-10} 
 &  & 0.80 & 0.038 & 106 &  & 0.1 & 50 & 0.04 & 100 \\ \cline{2-10} 
 & \multirow{2}{*}{0.75} & 0.75 & 0.05 & 103 &  & 0.12 & 50 & 0.052 & 99 \\ \cline{3-10} 
 &  & 0.85 & 0.04 & 156 &  & 0.106 & 76 & 0.04 & 152 \\ \cline{2-10} 
 & \multirow{2}{*}{0.80} & 0.80 & 0.048 & 147 &  & 0.116 & 78 & 0.052 & 155 \\ \cline{3-10} 
 &  & 0.90 & 0.04 & 232 &  & 0.092 & 120 & 0.042 & 240 \\ \hline
\end{tabular}
\vspace{-0.2em}
\begin{tablenotes}
      \small
      \item  \hspace{30pt} * When the fraction of information $\omega$ reaches 0.5.
\end{tablenotes}
\end{table}

\begin{table}[H]
\centering
\caption{Decision boundaries and sample sizes per dose arm for eROSE designs when $\delta=0.15$.} 
\label{tab:Bounderies_sample_size_eROSE_delta_0.15}
\setlength{\tabcolsep}{10pt}
\begin{tabular}{ccccc@{}ccccc}
\hline
\multirow{2}{*}{$p_H$} & \multirow{2}{*}{$\alpha_L$} & \multirow{2}{*}{$\alpha_H$} & \multicolumn{2}{c}{No interim} &  & \multicolumn{4}{c}{With one interim$^*$} \\ \cline{4-5} \cline{7-10} 
 &  &  & $\lambda$ & $n$ &  & $\lambda_1$ & $n_1$ & $\lambda$ & $n$ \\ \hline
\multirow{10}{*}{0.3} & \multirow{2}{*}{0.60} & 0.60 & 0 & 6 &  & 0.334 & 3 & 0 & 6 \\ \cline{3-10} 
 &  & 0.70 & 0.054 & 19 &  & 0.112 & 9 & 0.06 & 17 \\ \cline{2-10} 
 & \multirow{2}{*}{0.65} & 0.65 & 0.064 & 16 &  & 0.25 & 8 & 0.064 & 16 \\ \cline{3-10} 
 &  & 0.75 & 0.044 & 23 &  & 0.182 & 11 & 0.046 & 22 \\ \cline{2-10} 
 & \multirow{2}{*}{0.70} & 0.70 & 0.054 & 19 &  & 0.154 & 13 & 0.078 & 26 \\ \cline{3-10} 
 &  & 0.80 & 0.056 & 37 &  & 0.168 & 18 & 0.056 & 36 \\ \cline{2-10} 
 & \multirow{2}{*}{0.75} & 0.75 & 0.064 & 32 &  & 0.2 & 20 & 0.076 & 40 \\ \cline{3-10} 
 &  & 0.85 & 0.058 & 53 &  & 0.154 & 26 & 0.058 & 52 \\ \cline{2-10} 
 & \multirow{2}{*}{0.80} & 0.80 & 0.076 & 54 &  & 0.186 & 27 & 0.076 & 54 \\ \cline{3-10} 
 &  & 0.90 & 0.062 & 82 &  & 0.148 & 41 & 0.062 & 81 \\ \hline
\multirow{10}{*}{0.4} & \multirow{2}{*}{0.60} & 0.60 & 0 & 6 &  & 0.334 & 3 & 0 & 6 \\ \cline{3-10} 
 &  & 0.70 & 0.048 & 21 &  & 0.2 & 10 & 0.05 & 20 \\ \cline{2-10} 
 & \multirow{2}{*}{0.65} & 0.65 & 0.06 & 17 &  & 0.224 & 9 & 0.06 & 17 \\ \cline{3-10} 
 &  & 0.75 & 0.04 & 25 &  & 0.126 & 16 & 0.066 & 31 \\ \cline{2-10} 
 & \multirow{2}{*}{0.70} & 0.70 & 0.07 & 29 &  & 0.216 & 14 & 0.072 & 28 \\ \cline{3-10} 
 &  & 0.80 & 0.05 & 40 &  & 0.168 & 24 & 0.064 & 48 \\ \cline{2-10} 
 & \multirow{2}{*}{0.75} & 0.75 & 0.07 & 43 &  & 0.192 & 21 & 0.072 & 42 \\ \cline{3-10} 
 &  & 0.85 & 0.06 & 67 &  & 0.152 & 33 & 0.062 & 66 \\ \cline{2-10} 
 & \multirow{2}{*}{0.80} & 0.80 & 0.07 & 58 &  & 0.182 & 33 & 0.076 & 66 \\ \cline{3-10} 
 &  & 0.90 & 0.062 & 99 &  & 0.144 & 49 & 0.062 & 98 \\ \hline
\multirow{10}{*}{0.5} & \multirow{2}{*}{0.60} & 0.60 & 0 & 7 &  & 0.334 & 3 & 0 & 6 \\ \cline{3-10} 
 &  & 0.70 & 0.048 & 21 &  & 0.2 & 10 & 0.05 & 20 \\ \cline{2-10} 
 & \multirow{2}{*}{0.65} & 0.65 & 0.056 & 18 &  & 0.224 & 9 & 0.06 & 17 \\ \cline{3-10} 
 &  & 0.75 & 0.04 & 26 &  & 0.178 & 17 & 0.06 & 34 \\ \cline{2-10} 
 & \multirow{2}{*}{0.70} & 0.70 & 0.068 & 30 &  & 0.2 & 15 & 0.07 & 29 \\ \cline{3-10} 
 &  & 0.80 & 0.048 & 42 &  & 0.16 & 25 & 0.062 & 49 \\ \cline{2-10} 
 & \multirow{2}{*}{0.75} & 0.75 & 0.07 & 44 &  & 0.182 & 22 & 0.07 & 43 \\ \cline{3-10} 
 &  & 0.85 & 0.058 & 70 &  & 0.148 & 34 & 0.06 & 68 \\ \cline{2-10} 
 & \multirow{2}{*}{0.80} & 0.80 & 0.074 & 69 &  & 0.178 & 34 & 0.074 & 68 \\ \cline{3-10} 
 &  & 0.90 & 0.058 & 104 &  & 0.154 & 52 & 0.06 & 103 \\ \hline
\end{tabular}
\vspace{-0.2em}
\begin{tablenotes}
      \small
      \item  \hspace{30pt} * When the fraction of information $\omega$ reaches 0.5.
\end{tablenotes}
\end{table}

\newpage
\begin{table}[H]
\centering
\caption{Operating characteristics of eROSE design when $\delta =0.1$. The PCS, PET, and EN are exact values calculated from the binomial distribution, rather than being estimated through simulations.}
\label{tab:oc_eROSE_design_delta_0.1}
\begin{tabular}{cccccccccccc}
\hline
\multirow{3}{*}{$p_H$} & \multirow{3}{*}{$\alpha_L$} & \multirow{3}{*}{$\alpha_H$} & \multicolumn{2}{c}{No interim} &  & \multicolumn{6}{c}{With one interim$^*$} \\ \cline{4-5} \cline{7-12} 
 &  &   & \multicolumn{2}{c}{PCS} &  & \multicolumn{3}{c}{$p_L=0.3$} & \multicolumn{3}{c}{$p_L=0.2$} \\ \cline{4-5} \cline{7-12} 
 &   &  & $p_L=0.3$ & $p_L=0.2$ &  &PCS  &PET  &EN   &PCS  &PET  &EN  \\ \hline 
\multirow{10}{*}{0.3} & \multirow{2}{*}{0.60} & 0.60 & 0.69 & 0.61 &  & 0.64 & 0.23 & 16.9 & 0.61 & 0.40 & 15.4 \\ \cline{3-12} 
 &  & 0.70 & 0.65 & 0.70 &  & 0.63 & 0.17 & 28.5 & 0.70 & 0.36 & 25.7 \\ \cline{2-12} 
 & \multirow{2}{*}{0.65} & 0.65 & 0.67 & 0.66 &  & 0.69 & 0.18 & 31.9 & 0.65 & 0.39 & 28.3 \\ \cline{3-12} 
 &  & 0.75 & 0.70 & 0.75 &  & 0.67 & 0.15 & 49.1 & 0.75 & 0.40 & 42.6 \\ \cline{2-12} 
 & \multirow{2}{*}{0.70} & 0.70 & 0.71 & 0.70 &  & 0.74 & 0.10 & 55.2 & 0.71 & 0.31 & 48.9 \\ \cline{3-12} 
 &  & 0.80 & 0.72 & 0.80 &  & 0.70 & 0.10 & 80.9 & 0.82 & 0.38 & 69.0 \\ \cline{2-12} 
 & \multirow{2}{*}{0.75} & 0.75 & 0.78 & 0.75 &  & 0.76 & 0.09 & 76.4 & 0.75 & 0.35 & 66.1 \\ \cline{3-12} 
 &  & 0.85 & 0.78 & 0.85 &  & 0.75 & 0.10 & 115.9 & 0.85 & 0.47 & 93.5 \\ \cline{2-12} 
 & \multirow{2}{*}{0.80} & 0.80 & 0.82 & 0.80 &  & 0.80 & 0.07 & 115.0 & 0.80 & 0.38 & 96.9 \\ \cline{3-12} 
 &  & 0.90 & 0.81 & 0.90 &  & 0.81 & 0.07 & 183.7 & 0.90 & 0.50 & 142.5 \\ \hline
 &  &   & \multicolumn{2}{c}{PCS} &  & \multicolumn{3}{c}{$p_L=0.4$} & \multicolumn{3}{c}{$p_L=0.3$} \\ \cline{4-5} \cline{7-12} 
 &   &  & $p_L=0.4$ & $p_L=0.3$ &  &PCS  &PET  &EN   &PCS  &PET  &EN  \\ \hline  
\multirow{10}{*}{0.4} & \multirow{2}{*}{0.60} & 0.60 & 0.67 & 0.61 &  & 0.65 & 0.14 & 20.5 & 0.61 & 0.27 & 19.1 \\ \cline{3-12} 
 &  & 0.70 & 0.64 & 0.71 &  & 0.61 & 0.19 & 29.9 & 0.71 & 0.39 & 26.8 \\ \cline{2-12} 
 & \multirow{2}{*}{0.65} & 0.65 & 0.66 & 0.65 &  & 0.69 & 0.13 & 36.5 & 0.65 & 0.31 & 33.1 \\ \cline{3-12} 
 &  & 0.75 & 0.68 & 0.75 &  & 0.65 & 0.13 & 60.9 & 0.78 & 0.38 & 52.9 \\ \cline{2-12} 
 & \multirow{2}{*}{0.70} & 0.70 & 0.74 & 0.70 &  & 0.71 & 0.12 & 56.5 & 0.70 & 0.34 & 49.7 \\ \cline{3-12} 
 &  & 0.80 & 0.70 & 0.80 &  & 0.71 & 0.13 & 90.8 & 0.80 & 0.45 & 75.4 \\ \cline{2-12} 
 & \multirow{2}{*}{0.75} & 0.75 & 0.76 & 0.75 &  & 0.77 & 0.09 & 92.7 & 0.75 & 0.37 & 79.4 \\ \cline{3-12} 
 &  & 0.85 & 0.75 & 0.85 &  & 0.75 & 0.10 & 137.6 & 0.85 & 0.49 & 109.9 \\ \cline{2-12} 
 & \multirow{2}{*}{0.80} & 0.80 & 0.82 & 0.80 &  & 0.81 & 0.05 & 136.5 & 0.80 & 0.33 & 117.0 \\ \cline{3-12} 
 &  & 0.90 & 0.80 & 0.90 &  & 0.81 & 0.06 & 214.7 & 0.90 & 0.48 & 168.4 \\ \hline
 &  &   & \multicolumn{2}{c}{PCS} &  & \multicolumn{3}{c}{$p_L=0.5$} & \multicolumn{3}{c}{$p_L=0.4$} \\ \cline{4-5} \cline{7-12} 
 &   &  & $p_L=0.5$ & $p_L=0.4$ &  &PCS  &PET  &EN   &PCS  &PET  &EN  \\ \hline 
\multirow{10}{*}{0.5} & \multirow{2}{*}{0.60} & 0.60 & 0.67 & 0.61 &  & 0.65 & 0.14 & 20.4 & 0.61 & 0.28 & 19.0 \\ \cline{3-12} 
 &  & 0.70 & 0.63 & 0.70 &  & 0.60 & 0.20 & 29.9 & 0.70 & 0.39 & 26.7 \\ \cline{2-12} 
 & \multirow{2}{*}{0.65} & 0.65 & 0.65 & 0.65 &  & 0.68 & 0.13 & 37.3 & 0.66 & 0.32 & 33.6 \\ \cline{3-12} 
 &  & 0.75 & 0.67 & 0.75 &  & 0.65 & 0.13 & 57.2 & 0.76 & 0.36 & 50.2 \\ \cline{2-12} 
 & \multirow{2}{*}{0.70} & 0.70 & 0.73 & 0.70 &  & 0.71 & 0.13 & 57.2 & 0.71 & 0.36 & 50.2 \\ \cline{3-12} 
 &  & 0.80 & 0.73 & 0.80 &  & 0.70 & 0.14 & 93.2 & 0.80 & 0.46 & 76.9 \\ \cline{2-12} 
 & \multirow{2}{*}{0.75} & 0.75 & 0.78 & 0.75 &  & 0.76 & 0.10 & 94.3 & 0.75 & 0.38 & 80.3 \\ \cline{3-12} 
 &  & 0.85 & 0.77 & 0.85 &  & 0.75 & 0.08 & 145.6 & 0.85 & 0.44 & 118.4 \\ \cline{2-12} 
 & \multirow{2}{*}{0.80} & 0.80 & 0.81 & 0.80 &  & 0.81 & 0.06 & 150.1 & 0.80 & 0.39 & 124.8 \\ \cline{3-12} 
 &  & 0.90 & 0.81 & 0.90 &  & 0.81 & 0.07 & 231.8 & 0.90 & 0.53 & 176.8 \\ \hline
\end{tabular}
\vspace{-0.2em}
\begin{tablenotes}
      \small
      \item  \hspace{30pt} * When the fraction of information $\omega$ reaches 0.5.
\end{tablenotes}
\end{table}

\begin{table}[H]
\centering
\caption{Operating characteristics of eROSE design when $\delta =0.15$. The PCS, PET, and EN are exact values calculated from the binomial distribution, rather than being estimated through simulations.}
\label{tab:oc_eROSE_design_delta_0.15}
\begin{tabular}{cccccccccccc}
\hline
\multirow{3}{*}{$p_H$} & \multirow{3}{*}{$\alpha_L$} & \multirow{3}{*}{$\alpha_H$} & \multicolumn{2}{c}{No interim} &  & \multicolumn{6}{c}{With one interim$^*$} \\ \cline{4-5} \cline{7-12} 
 &  &   & \multicolumn{2}{c}{PCS} &  & \multicolumn{3}{c}{$p_L=0.3$} & \multicolumn{3}{c}{$p_L=0.15$} \\ \cline{4-5} \cline{7-12} 
 &   &  & $p_L=0.3$ & $p_L=0.15$ &  &PCS  &PET  &EN   &PCS  &PET  &EN  \\ \hline 
\multirow{10}{*}{0.3} & \multirow{2}{*}{0.60} & 0.60 & 0.62 & 0.61 &  & 0.62 & 0.09 & 5.7 & 0.62 & 0.14 & 5.6 \\ \cline{3-12} 
 &  & 0.70 & 0.70 & 0.70 &  & 0.65 & 0.22 & 15.2 & 0.71 & 0.47 & 13.3 \\ \cline{2-12} 
 & \multirow{2}{*}{0.65} & 0.65 & 0.72 & 0.65 &  & 0.71 & 0.08 & 15.3 & 0.66 & 0.21 & 14.3 \\ \cline{3-12} 
 &  & 0.75 & 0.69 & 0.76 &  & 0.67 & 0.12 & 20.7 & 0.76 & 0.33 & 18.4 \\ \cline{2-12} 
 & \multirow{2}{*}{0.70} & 0.70 & 0.70 & 0.70 &  & 0.73 & 0.14 & 24.2 & 0.71 & 0.40 & 20.9 \\ \cline{3-12} 
 &  & 0.80 & 0.74 & 0.81 &  & 0.72 & 0.10 & 34.2 & 0.81 & 0.37 & 29.3 \\ \cline{2-12} 
 & \multirow{2}{*}{0.75} & 0.75 & 0.75 & 0.76 &  & 0.79 & 0.06 & 38.8 & 0.76 & 0.28 & 34.4 \\ \cline{3-12} 
 &  & 0.85 & 0.77 & 0.85 &  & 0.75 & 0.09 & 49.8 & 0.85 & 0.42 & 41.1 \\ \cline{2-12} 
 & \multirow{2}{*}{0.80} & 0.80 & 0.83 & 0.80 &  & 0.82 & 0.05 & 52.6 & 0.81 & 0.32 & 45.5 \\ \cline{3-12} 
 &  & 0.90 & 0.83 & 0.90 &  & 0.81 & 0.06 & 78.7 & 0.90 & 0.46 & 62.5 \\ \hline
 &  &   & \multicolumn{2}{c}{PCS} &  & \multicolumn{3}{c}{$p_L=0.4$} & \multicolumn{3}{c}{$p_L=0.25$} \\ \cline{4-5} \cline{7-12} 
 &   &  & $p_L=0.4$ & $p_L=0.25$ &  &PCS  &PET  &EN   &PCS  &PET  &EN  \\ \hline 
\multirow{10}{*}{0.4} & \multirow{2}{*}{0.60} & 0.60 & 0.62 & 0.60 &  & 0.61 & 0.10 & 5.7 & 0.61 & 0.18 & 5.5 \\ \cline{3-12} 
 &  & 0.70 & 0.68 & 0.71 &  & 0.66 & 0.13 & 18.7 & 0.71 & 0.32 & 16.8 \\ \cline{2-12} 
 & \multirow{2}{*}{0.65} & 0.65 & 0.70 & 0.65 &  & 0.68 & 0.11 & 16.1 & 0.67 & 0.28 & 14.8 \\ \cline{3-12} 
 &  & 0.75 & 0.67 & 0.76 &  & 0.69 & 0.18 & 28.2 & 0.76 & 0.49 & 23.7 \\ \cline{2-12} 
 & \multirow{2}{*}{0.70} & 0.70 & 0.75 & 0.70 &  & 0.73 & 0.09 & 26.8 & 0.70 & 0.29 & 24.0 \\ \cline{3-12} 
 &  & 0.80 & 0.72 & 0.80 &  & 0.75 & 0.09 & 45.8 & 0.80 & 0.39 & 38.6 \\ \cline{2-12} 
 & \multirow{2}{*}{0.75} & 0.75 & 0.78 & 0.76 &  & 0.76 & 0.08 & 40.4 & 0.76 & 0.33 & 35.1 \\ \cline{3-12} 
 &  & 0.85 & 0.79 & 0.85 &  & 0.77 & 0.08 & 63.3 & 0.85 & 0.44 & 51.4 \\ \cline{2-12} 
 & \multirow{2}{*}{0.80} & 0.80 & 0.80 & 0.80 &  & 0.82 & 0.05 & 64.3 & 0.80 & 0.34 & 54.7 \\ \cline{3-12} 
 &  & 0.90 & 0.83 & 0.90 &  & 0.81 & 0.06 & 95.0 & 0.90 & 0.49 & 74.1 \\ \hline
 &  &   & \multicolumn{2}{c}{PCS} &  & \multicolumn{3}{c}{$p_L=0.5$} & \multicolumn{3}{c}{$p_L=0.35$} \\ \cline{4-5} \cline{7-12} 
 &   &  & $p_L=0.5$ & $p_L=0.35$ &  &PCS  &PET  &EN   &PCS  &PET  &EN  \\ \hline 
\multirow{10}{*}{0.5} & \multirow{2}{*}{0.60} & 0.60 & 0.60 & 0.62 &  & 0.60 & 0.11 & 5.7 & 0.61 & 0.19 & 5.4 \\ \cline{3-12} 
 &  & 0.70 & 0.68 & 0.70 &  & 0.66 & 0.13 & 18.7 & 0.70 & 0.33 & 16.7 \\ \cline{2-12} 
 & \multirow{2}{*}{0.65} & 0.65 & 0.69 & 0.66 &  & 0.68 & 0.12 & 16.0 & 0.66 & 0.29 & 14.7 \\ \cline{3-12} 
 &  & 0.75 & 0.66 & 0.75 &  & 0.70 & 0.11 & 32.0 & 0.76 & 0.37 & 27.7 \\ \cline{2-12} 
 & \multirow{2}{*}{0.70} & 0.70 & 0.74 & 0.70 &  & 0.72 & 0.10 & 27.6 & 0.71 & 0.32 & 24.5 \\ \cline{3-12} 
 &  & 0.80 & 0.71 & 0.80 &  & 0.74 & 0.10 & 46.6 & 0.80 & 0.42 & 39.0 \\ \cline{2-12} 
 & \multirow{2}{*}{0.75} & 0.75 & 0.77 & 0.75 &  & 0.75 & 0.09 & 41.2 & 0.76 & 0.36 & 35.5 \\ \cline{3-12} 
 &  & 0.85 & 0.78 & 0.85 &  & 0.76 & 0.09 & 64.9 & 0.85 & 0.46 & 52.3 \\ \cline{2-12} 
 & \multirow{2}{*}{0.80} & 0.80 & 0.83 & 0.80 &  & 0.81 & 0.06 & 66.1 & 0.80 & 0.37 & 55.5 \\ \cline{3-12} 
 &  & 0.90 & 0.82 & 0.90 &  & 0.81 & 0.05 & 100.6 & 0.90 & 0.45 & 80.3 \\ \hline
\end{tabular}
\vspace{-0.2em}
\begin{tablenotes}
      \small
      \item  \hspace{30pt} * When the fraction of information $\omega$ reaches 0.5.
\end{tablenotes}
\end{table}

\begin{table}[H]
\centering
\caption{Decision boundaries and global minimum sample size based on the exact binomial distribution for the two-stage design, with $\omega=0.5$.} 
\label{tab:Bounderies_sample_size_eROSE_global_min}
\setlength{\tabcolsep}{10pt}
\begin{tabular}{ccccccc@{}ccccc}
\hline
\multirow{2}{*}{$p_H$} & \multirow{2}{*}{$\alpha_L$} & \multirow{2}{*}{$\alpha_H$} & \multicolumn{4}{c}{$\delta = 0.1$} &  & \multicolumn{4}{c}{$\delta = 0.15$} \\ \cline{4-7} \cline{9-12} 
 &  &  & $\lambda_1$ & $n_1$ & $\lambda$ & $n$ &  & $\lambda_1$ & $n_1$ & $\lambda$ & $n$ \\ \hline
\multirow{10}{*}{0.3} & \multirow{2}{*}{0.60} & 0.60 & 0.1 & 10 & 0.054 & 19 &  & 0.334 & 3 & 0 & 6 \\ \cline{3-12} 
 &  & 0.70 & 0.126 & 16 & 0.034 & 31 &  & 0.112 & 9 & 0.06 & 17 \\ \cline{2-12} 
 & \multirow{2}{*}{0.65} & 0.65 & 0.216 & 14 & 0.038 & 27 &  & 0.126 & 8 & 0.068 & 15 \\ \cline{3-12} 
 &  & 0.75 & 0.112 & 27 & 0.038 & 53 &  & 0.182 & 11 & 0.046 & 22 \\ \cline{2-12} 
 & \multirow{2}{*}{0.70} & 0.70 & 0.174 & 23 & 0.044 & 46 &  & 0.4 & 10 & 0.054 & 19 \\ \cline{3-12} 
 &  & 0.80 & 0.126 & 40 & 0.038 & 80 &  & 0.168 & 18 & 0.056 & 36 \\ \cline{2-12} 
 & \multirow{2}{*}{0.75} & 0.75 & 0.126 & 40 & 0.05 & 80 &  & 0.314 & 16 & 0.064 & 32 \\ \cline{3-12} 
 &  & 0.85 & 0.1 & 61 & 0.042 & 122 &  & 0.154 & 26 & 0.058 & 52 \\ \cline{2-12} 
 & \multirow{2}{*}{0.80} & 0.80 & 0.118 & 60 & 0.052 & 119 &  & 0.15 & 27 & 0.076 & 53 \\ \cline{3-12} 
 &  & 0.90 & 0.124 & 90 & 0.04 & 179 &  & 0.148 & 41 & 0.062 & 81 \\ \hline
\multirow{10}{*}{0.4} & \multirow{2}{*}{0.60} & 0.60 & 0.112 & 9 & 0.056 & 18 &  & 0.334 & 3 & 0 & 6 \\ \cline{3-12} 
 &  & 0.70 & 0.118 & 17 & 0.032 & 33 &  & 0.112 & 9 & 0.06 & 17 \\ \cline{2-12} 
 & \multirow{2}{*}{0.65} & 0.65 & 0.268 & 15 & 0.036 & 29 &  & 0.224 & 9 & 0.06 & 17 \\ \cline{3-12} 
 &  & 0.75 & 0.138 & 29 & 0.036 & 58 &  & 0.232 & 13 & 0.04 & 25 \\ \cline{2-12} 
 & \multirow{2}{*}{0.70} & 0.70 & 0.134 & 30 & 0.05 & 60 &  & 0.144 & 14 & 0.076 & 27 \\ \cline{3-12} 
 &  & 0.80 & 0.182 & 44 & 0.036 & 88 &  & 0.2 & 20 & 0.05 & 40 \\ \cline{2-12} 
 & \multirow{2}{*}{0.75} & 0.75 & 0.182 & 44 & 0.046 & 87 &  & 0.192 & 21 & 0.072 & 42 \\ \cline{3-12} 
 &  & 0.85 & 0.162 & 68 & 0.038 & 136 &  & 0.126 & 32 & 0.064 & 64 \\ \cline{2-12} 
 & \multirow{2}{*}{0.80} & 0.80 & 0.116 & 70 & 0.052 & 139 &  & 0.242 & 29 & 0.07 & 58 \\ \cline{3-12} 
 &  & 0.90 & 0.154 & 105 & 0.04 & 210 &  & 0.126 & 48 & 0.064 & 96 \\ \hline
\multirow{10}{*}{0.5} & \multirow{2}{*}{0.60} & 0.60 & 0.112 & 9 & 0.056 & 18 &  & 0.334 & 3 & 0 & 6 \\ \cline{3-12} 
 &  & 0.70 & 0.118 & 17 & 0.032 & 33 &  & 0.112 & 9 & 0.056 & 18 \\ \cline{2-12} 
 & \multirow{2}{*}{0.65} & 0.65 & 0.4 & 15 & 0.034 & 30 &  & 0.224 & 9 & 0.06 & 17 \\ \cline{3-12} 
 &  & 0.75 & 0.134 & 30 & 0.034 & 60 &  & 0.232 & 13 & 0.04 & 26 \\ \cline{2-12} 
 & \multirow{2}{*}{0.70} & 0.70 & 0.13 & 31 & 0.05 & 61 &  & 0.2 & 15 & 0.07 & 29 \\ \cline{3-12} 
 &  & 0.80 & 0.1 & 50 & 0.04 & 100 &  & 0.24 & 21 & 0.048 & 42 \\ \cline{2-12} 
 & \multirow{2}{*}{0.75} & 0.75 & 0.12 & 50 & 0.052 & 99 &  & 0.182 & 22 & 0.07 & 43 \\ \cline{3-12} 
 &  & 0.85 & 0.106 & 76 & 0.04 & 152 &  & 0.148 & 34 & 0.06 & 68 \\ \cline{2-12} 
 & \multirow{2}{*}{0.80} & 0.80 & 0.138 & 73 & 0.05 & 145 &  & 0.152 & 33 & 0.076 & 66 \\ \cline{3-12} 
 &  & 0.90 & 0.08 & 115 & 0.044 & 229 &  & 0.154 & 52 & 0.06 & 103 \\ \hline
\end{tabular}
\vspace{-0.2em}
\begin{tablenotes}
      \small
      \item  \hspace{30pt} * When the fraction of information $\omega$ reaches 0.5.
\end{tablenotes}
\end{table}


\begin{table}[htbp]
\centering
\caption{Operating characteristics of the two-stage design, based on decision boundaries and global minimum sample size under the exact binomial distribution, with $\omega=0.5$. The PCS, PET, and EN are exact values calculated from the binomial distribution, rather than being estimated through simulations.}
\label{tab:sim_result_design_interim_global_min}
 \setlength{\tabcolsep}{4.5pt}
\begin{tabular}{cccccccccc@{}cccccc}
\hline
\multirow{3}{*}{$p_H$} & \multirow{3}{*}{{$\alpha_L$}} & \multirow{3}{*}{{$\alpha_H$}} & \multicolumn{6}{c}{$\delta = 0.1$} &  & \multicolumn{6}{c}{$\delta = 0.15$} \\ \cline{4-9} \cline{11-16}  
&  &  & \multicolumn{3}{c}{$p_L=0.3$} & \multicolumn{3}{c}{$p_L=0.2$} &  & \multicolumn{3}{c}{$p_L=0.3$} & \multicolumn{3}{c}{$p_L=0.15$} \\ \cline{4-9} \cline{11-16} 
 &  &  & PCS & PET & EN & PCS & PET & EN &  & PCS & PET & EN & PCS & PET & EN \\ \hline
\multirow{10}{*}{0.3} & \multirow{2}{*}{0.60} & 0.60 & 0.64 & 0.23 & 16.9 & 0.61 & 0.40 & 15.4 &  & 0.62 & 0.09 & 5.7 & 0.62 & 0.14 & 5.6 \\ \cline{3-16} 
 &  & 0.70 & 0.63 & 0.17 & 28.5 & 0.70 & 0.36 & 25.7 &  & 0.65 & 0.22 & 15.2 & 0.71 & 0.47 & 13.3 \\ \cline{2-16} 
 & \multirow{2}{*}{0.65} & 0.65 & 0.66 & 0.07 & 26.0 & 0.65 & 0.18 & 24.7 &  & 0.67 & 0.21 & 13.6 & 0.67 & 0.43 & 12.0 \\ \cline{3-16} 
 &  & 0.75 & 0.67 & 0.15 & 49.1 & 0.75 & 0.40 & 42.6 &  & 0.67 & 0.12 & 20.7 & 0.76 & 0.33 & 18.4 \\ \cline{2-16} 
 & \multirow{2}{*}{0.70} & 0.70 & 0.70 & 0.07 & 44.3 & 0.70 & 0.22 & 40.8 &  & 0.70 & 0.01 & 18.9 & 0.70 & 0.05 & 18.6 \\ \cline{3-16} 
 &  & 0.80 & 0.71 & 0.09 & 76.4 & 0.80 & 0.35 & 66.1 &  & 0.72 & 0.10 & 34.2 & 0.81 & 0.37 & 29.3 \\ \cline{2-16} 
 & \multirow{2}{*}{0.75} & 0.75 & 0.76 & 0.09 & 76.4 & 0.75 & 0.35 & 66.1 &  & 0.75 & 0.02 & 31.7 & 0.76 & 0.09 & 30.6 \\ \cline{3-16} 
 &  & 0.85 & 0.75 & 0.10 & 115.9 & 0.85 & 0.47 & 93.5 &  & 0.75 & 0.09 & 49.8 & 0.85 & 0.42 & 41.1 \\ \cline{2-16} 
 & \multirow{2}{*}{0.80} & 0.80 & 0.80 & 0.07 & 115.0 & 0.80 & 0.38 & 96.9 &  & 0.80 & 0.09 & 50.7 & 0.81 & 0.44 & 41.5 \\ \cline{3-16} 
 &  & 0.90 & 0.80 & 0.03 & 176.3 & 0.90 & 0.33 & 149.4 &  & 0.81 & 0.06 & 78.7 & 0.90 & 0.46 & 62.5 \\ \hline
\multicolumn{1}{l}{} & \multicolumn{1}{l}{} &  & \multicolumn{3}{c}{$p_L=0.4$} & \multicolumn{3}{c}{$p_L=0.3$} &  & \multicolumn{3}{c}{$p_L=0.4$} & \multicolumn{3}{c}{$p_L=0.25$} \\ \cline{4-9} \cline{11-16} 
\multicolumn{1}{l}{} & \multicolumn{1}{l}{} & \multicolumn{1}{l}{} & PCS & PET & EN & PCS & PET & EN &  & PCS & PET & EN & PCS & PET & EN \\ \hline
\multirow{10}{*}{0.4} & \multirow{2}{*}{0.60} & 0.60 & 0.63 & 0.24 & 15.9 & 0.60 & 0.39 & 14.5 &  & 0.61 & 0.10 & 5.7 & 0.61 & 0.18 & 5.5 \\ \cline{3-16} 
 &  & 0.70 & 0.61 & 0.19 & 29.9 & 0.71 & 0.39 & 26.8 &  & 0.63 & 0.24 & 15.1 & 0.70 & 0.47 & 13.2 \\ \cline{2-16} 
 & \multirow{2}{*}{0.65} & 0.65 & 0.65 & 0.05 & 28.4 & 0.65 & 0.12 & 27.3 &  & 0.68 & 0.11 & 16.1 & 0.67 & 0.28 & 14.8 \\ \cline{3-16} 
 &  & 0.75 & 0.66 & 0.11 & 54.7 & 0.75 & 0.33 & 48.4 &  & 0.66 & 0.08 & 24.0 & 0.76 & 0.26 & 21.9 \\ \cline{2-16} 
 & \multirow{2}{*}{0.70} & 0.70 & 0.71 & 0.12 & 56.5 & 0.70 & 0.34 & 49.7 &  & 0.70 & 0.17 & 24.8 & 0.71 & 0.44 & 21.3 \\ \cline{3-16} 
 &  & 0.80 & 0.70 & 0.03 & 86.6 & 0.80 & 0.18 & 80.1 &  & 0.70 & 0.07 & 38.5 & 0.81 & 0.31 & 33.9 \\ \cline{2-16} 
 & \multirow{2}{*}{0.75} & 0.75 & 0.75 & 0.03 & 85.6 & 0.75 & 0.18 & 79.3 &  & 0.76 & 0.08 & 40.4 & 0.76 & 0.33 & 35.1 \\ \cline{3-16} 
 &  & 0.85 & 0.75 & 0.02 & 134.5 & 0.85 & 0.20 & 122.5 &  & 0.75 & 0.13 & 60.0 & 0.85 & 0.53 & 46.9 \\ \cline{2-16} 
 & \multirow{2}{*}{0.80} & 0.80 & 0.80 & 0.07 & 134.1 & 0.80 & 0.40 & 111.7 &  & 0.80 & 0.02 & 57.4 & 0.80 & 0.19 & 52.6 \\ \cline{3-16} 
 &  & 0.90 & 0.80 & 0.01 & 209.0 & 0.90 & 0.19 & 189.9 &  & 0.80 & 0.09 & 91.8 & 0.90 & 0.56 & 69.0 \\ \hline
 &  &  & \multicolumn{3}{c}{$p_L=0.5$} & \multicolumn{3}{c}{$p_L=0.4$} &  & \multicolumn{3}{c}{$p_L=0.5$} & \multicolumn{3}{c}{$p_L=0.35$} \\ \cline{4-9} \cline{11-16} 
\multicolumn{1}{l}{} & \multicolumn{1}{l}{} & \multicolumn{1}{l}{} & PCS & PET & EN & PCS & PET & EN &  & PCS & PET & EN & PCS & PET & EN \\ \hline
\multirow{10}{*}{0.5} & \multirow{2}{*}{0.60} & 0.60 & 0.62 & 0.24 & 15.8 & 0.60 & 0.39 & 14.5 &  & 0.60 & 0.11 & 5.7 & 0.61 & 0.19 & 5.4 \\ \cline{3-16} 
 &  & 0.70 & 0.60 & 0.20 & 29.9 & 0.70 & 0.39 & 26.7 &  & 0.62 & 0.24 & 15.8 & 0.71 & 0.48 & 13.7 \\ \cline{2-16} 
 & \multirow{2}{*}{0.65} & 0.65 & 0.65 & 0.01 & 29.9 & 0.65 & 0.03 & 29.5 &  & 0.68 & 0.12 & 16.0 & 0.66 & 0.29 & 14.7 \\ \cline{3-16} 
 &  & 0.75 & 0.65 & 0.12 & 56.3 & 0.75 & 0.35 & 49.5 &  & 0.65 & 0.08 & 24.9 & 0.76 & 0.27 & 22.5 \\ \cline{2-16} 
 & \multirow{2}{*}{0.70} & 0.70 & 0.71 & 0.13 & 57.2 & 0.71 & 0.36 & 50.2 &  & 0.72 & 0.10 & 27.6 & 0.71 & 0.32 & 24.5 \\ \cline{3-16} 
 &  & 0.80 & 0.70 & 0.14 & 93.2 & 0.80 & 0.46 & 76.9 &  & 0.70 & 0.04 & 41.1 & 0.80 & 0.23 & 37.2 \\ \cline{2-16} 
 & \multirow{2}{*}{0.75} & 0.75 & 0.76 & 0.10 & 94.3 & 0.75 & 0.38 & 80.3 &  & 0.75 & 0.09 & 41.2 & 0.76 & 0.36 & 35.5 \\ \cline{3-16} 
 &  & 0.85 & 0.75 & 0.08 & 145.6 & 0.85 & 0.44 & 118.4 &  & 0.76 & 0.09 & 64.9 & 0.85 & 0.46 & 52.3 \\ \cline{2-16} 
 & \multirow{2}{*}{0.80} & 0.80 & 0.80 & 0.04 & 142.1 & 0.80 & 0.30 & 123.6 &  & 0.80 & 0.09 & 63.1 & 0.80 & 0.45 & 51.2 \\ \cline{3-16} 
 &  & 0.90 & 0.80 & 0.10 & 217.5 & 0.90 & 0.61 & 159.4 &  & 0.81 & 0.05 & 100.6 & 0.90 & 0.45 & 80.3 \\ \hline
\end{tabular}
\end{table}

\newpage

\section{Simulation Setting for U-MET design}\label{supsec:U-MET}
Following BOIN12 design \citep{BOIN12}, U-MET design \citep{DAngelo2024sup} utilizes utility to quantify the response-toxicity trade-off. Specifically, with binary response and toxicity endpoints, there are four possible response-toxicity outcome combinations: 1 (response, no toxicity), 2 (no response, no toxicity), 3 (response, toxicity), and 4 (no response, toxicity). The utility scores ($u_1$,$u_2$, $u_3$, $u_4$) are elicited from clinicians to reflect the desirability of these four possible outcomes). A higher value of $u_k$, $k=1,\dots, 4$ indicates a more desirable outcome. The recommended utility scoring convention assigns the most desirable outcome (response, no toxicity) a score of 100, i.e.,$u_1=100$, and the least desirable outcome (no response, toxicity) as s score of 0, i.e., $u_4=0$. Intermediate values between 0 and 100 for $u_2$ and $u_3$, reflecting their relative desirability. An example is provided in \ref{tab:utility}. 
\begin{table}[!h]
    \centering
      \renewcommand{\arraystretch}{1} 
    \setlength{\tabcolsep}{12pt} 
    \caption{Example of utility table}
    \begin{tabular}{ccc}
        \hline
        \multirow{2}{*}{Toxicity} & \multicolumn{2}{c}{Response} \\
        \cline{2-3}
         & Yes & No \\
        \hline
        No  & $u_1 = 100$ & $u_2 = 35$ \\
        Yes & $u_3 = 65$ & $u_4 = 0$ \\
        \hline
    \end{tabular}
    \label{tab:utility}
\end{table}

The desirability of dose $d_j$ is computed as $u(d_j) = \sum_{k=1}^4\Psi_ku_k,$ where $\Psi_k$ are probabilities of outcome $k$. Let $n_j$ denote the number of patients treated at dose $d_j$ and $X_{j,k}$ the number of patients experiencing outcome $k$ among those $n_j$ patients. Following \citet{BOIN12}, the standardized mean utility $u^*(d_j)$ is defined as $u^*(d_j) = u(d_j)/100 \in [0,1]$. It is modeled by a binomial distribution with “quasi-binomial” data ($x(d_j), n_j$) as
$$x(d_j)|u^*(d_j) \sim Binomial(n_j,u^*(d_j)),$$
where $x(d_j) = \sum_{k=1}^4 u_k^{*}X_{j,k}$ represents the number of ``quasi events"among $n_j$ patients. 

The U-MET design selects doses based on a hypothesis framework, incorporating both frequentist and Bayesian approaches. Simulation studies by \citet{DAngelo2024sup} indicate comparable operating characteristics between the two versions; thus, we focus on the frequentist approach for comparison. Define the utility difference as $u^*_{\Delta} = u^*(d_H) -u^*(d_L)$ and consider the following hypotheses: 
$$H_0: u^*_{\Delta} \leq 0 \text{ versus } H_1: u^*_{\Delta}>0.$$
The hypothesis is evaluated using the one-sided confidence interval (CI) of $\widehat{u}^*_{\Delta}$, given by ($LB(\gamma), \infty$), where
$$LB(\gamma) = \widehat{u}^*_{\Delta} - z_{1-\gamma}\sqrt{\widehat{\tau}_L^2 + \widehat{\tau}_H^2},$$
with $\widehat{u}^*_{\Delta} = \widehat{u}^*(d_H) - \widehat{u}^*(d_L)$, $\widehat{u}^*(d_j) =x(d_j)/n_j$, and $\widehat{\tau}_j = \{\widehat{u}^*(d_j)(1-\widehat{u}^*(d_j))\}/n_j$ for $j=L,H$. Here, $z_{1-\gamma}$ denotes the ($1-\gamma$) quantile of the standard normal distribution, computed as $\Phi^{-1}(1-\lambda)$.

Let $\gamma_H$ and $\gamma_L$ denote prespecified significance levels, with $\gamma_L > \gamma_H$. The frequentist U-MET design selects doses based on the following rules:
\begin{itemize}
 \setlength{\itemsep}{0pt}
 \setlength{\parskip}{0pt}  
    \item Select the $d_H$ as the OBD if $LB(\gamma_H) >0$;
    \item Select the $d_L$ as the  OBD if $LB(\gamma_L) <0$;
    \item Otherwise, i.e., $LB(\gamma_H) <0$ and $LB(\gamma_L) >0$, make a ``consider high dose" decision.
\end{itemize}
Since U-MET allows three possible decisions, ``Select $d_L$", ``Select $d_H$", or ``Consider high dose", we adapt the classification to make the comparison relatively fairer. Specifically, when the decision is ``Consider high dose," we classify it as ``Select $d_H$" in our simulation study.    

For our simulations, we used the recommended parameter settings for $\gamma_j$, i.e., $\gamma_H = 0.2$ and $\gamma_L=0.34$. We used the utility scores listed in \ref{tab:utility}, consistent with those in \citep{DAngelo2024sup}. The true toxicity rates of $d_H$ and $d_L$ were fixed at 0.25 and 0.3, respectively, across all parameter configurations. Each design was evaluated using 10,000 simulations.  

\newpage

\section{\texorpdfstring{Sensitivity Analysis When $p_H$ deviates}{Sensitivity Analysis When pH deviates}}\label{sec:sens_p_deviate}

To design the trial, we need to pre-specify the expected response rate at the high dose, denoted as $p_H$,based on results from a phase I dose-escalation trial or pre-clinical data. In practice, the true response rate at the high dose, denoted as $p_H^{true}$, may differ from the pre-specified $p_H$. We evaluate the operating characteristics of the ROSE design when $p_H^{true}$ deviates from the assumed $p_H$. 

\begin{figure}[H]
    \centering
    \includegraphics[width=1\textwidth]{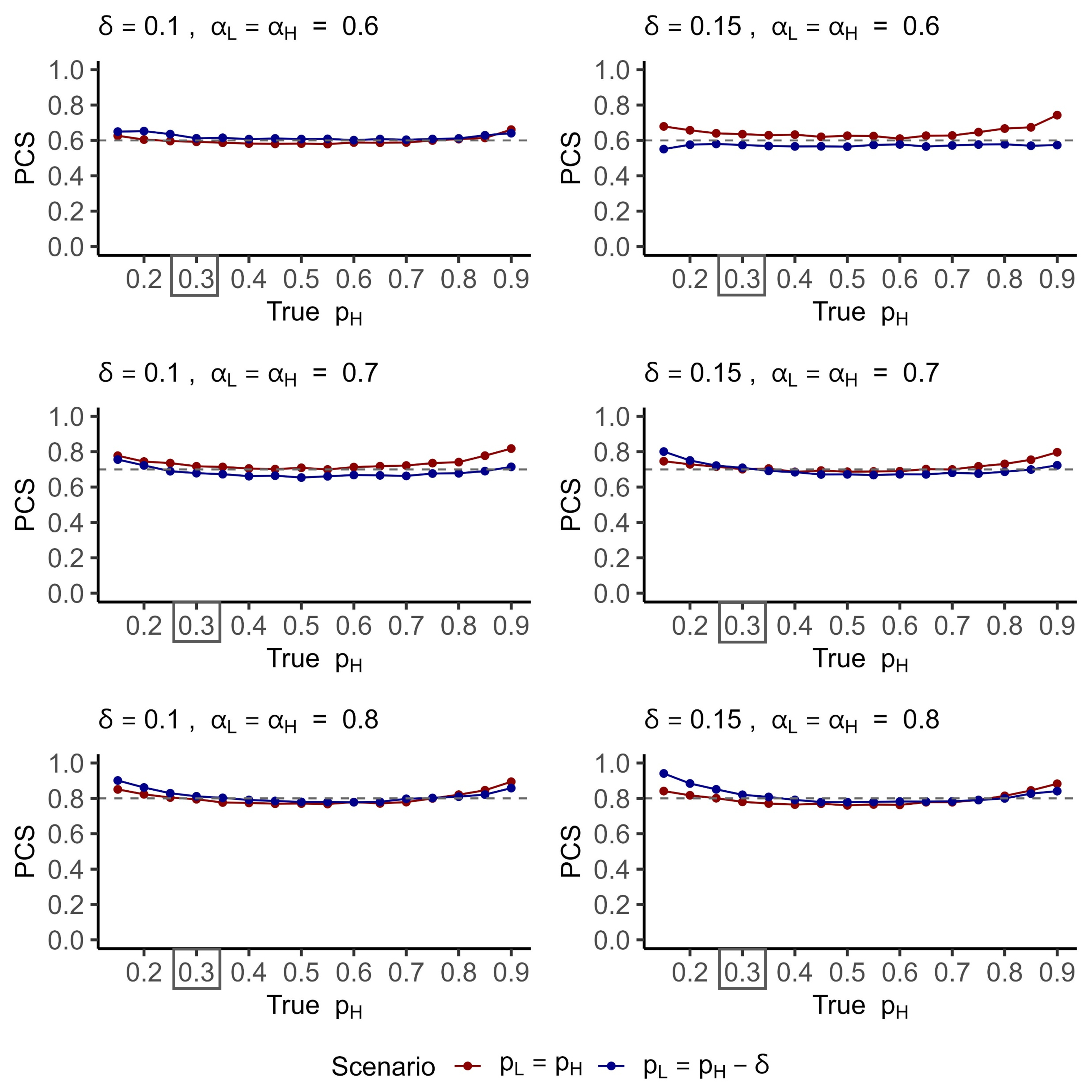}
    \caption{Operating characteristics of the ROSE design when true $p_H$ deviates from the anticipated $p_H$ used to design the trial. The trials are designed assuming $p_H = 0.3$ (indicated by the square on the x-axis). The dashed horizontal line represents the pre-specified desirable levels.}
    \label{fig:pH-deviate_0.3}
\end{figure}

Figure \ref{fig:pH-deviate_0.3} presents the PCSs under both scenarios of interest ($\mathcal{S}_L$ and $\mathcal{S}_H$) when the trial is designed assuming $p_H=0.3$, across different values of $\delta$ and target levels $\alpha_L$ and $\alpha_H$. Even when $p_H^{true}$ deviates from the assumed $p_H$, the PCS remains close to the pre-specified desirable levels.

\section{\texorpdfstring{Sensitivity Analysis When $n$ deviates}{Sensitivity Analysis When n deviates}} \label{sec:sens_n_deviate}

In practice, the sample size enrolled in each dose arm may not exactly match the planned number. In this section, we present the operating characteristics of the ROSE design when the sample sizes in one or both dose arms deviate from the planned sample size $n$. We consider four types of deviations: (1) $n_L=n$, while $n_H$ deviates from $n$; (2) $n_H=n$, while $n_L$ deviates from $n$, (3) Both $n_L$ and $n_H$ deviates in the same direction, meaning both increase or both decrease simultaneously, with $n_L=n_H \neq n$; (4) Both $n_L$ and $n_H$ deviates in opposite directions, such that when $n_L$ decreases, $n_H$ increases, maintaining a constant total sample size of  $n_L + n_H =2n$. 

\begin{figure}[H]
    \centering
    \includegraphics[width=1\textwidth]{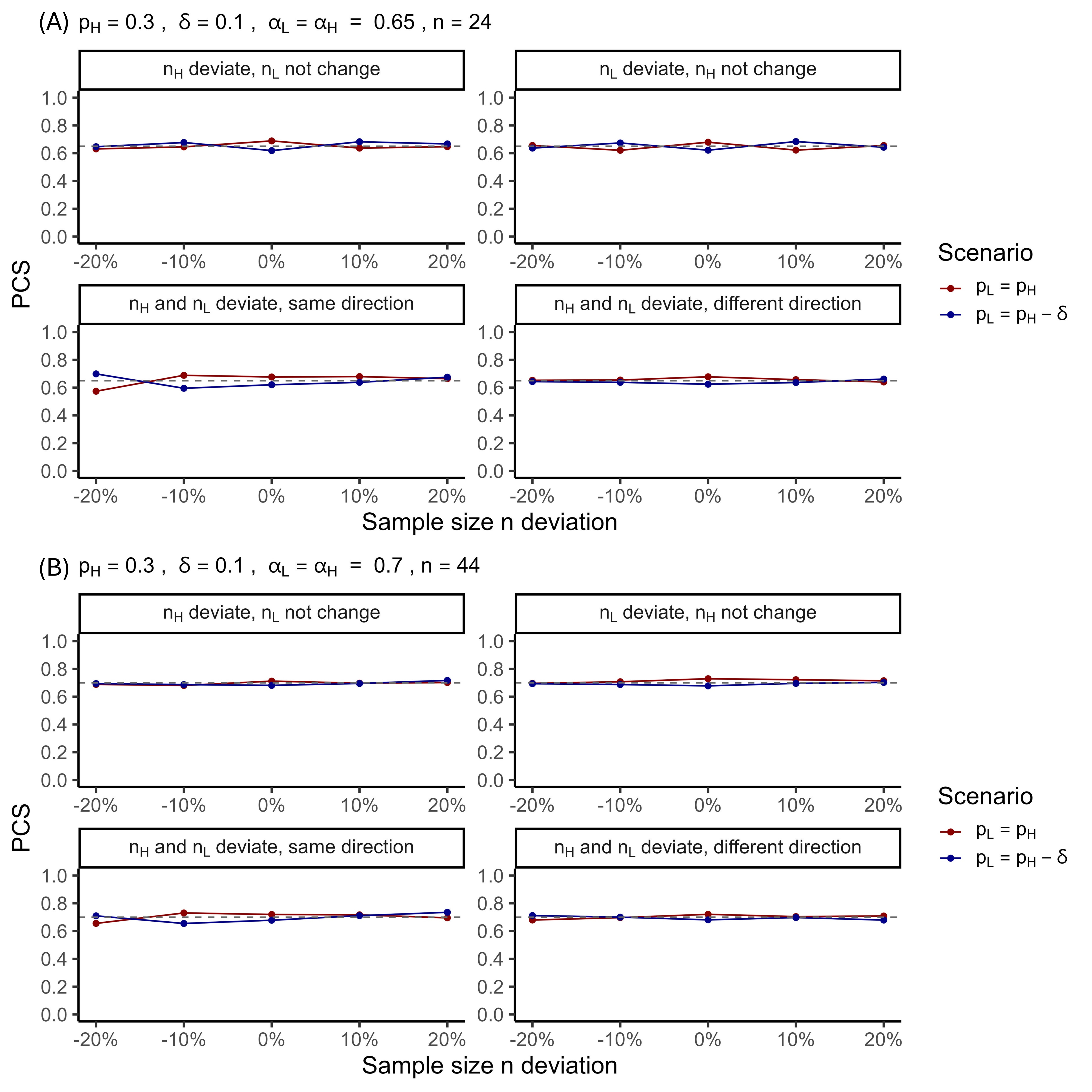}
    \caption{Operating characteristics of the ROSE design when enrolled sample sizes $n_L$ and/or $n_H$ deviate from the planned $n$ per dose arm. The trials are designed assuming $p_H = 0.3$ and $\delta=0.1$. The dashed horizontal line represents the pre-specified desirable levels.}
    \label{fig:samplesize-deviate_pH_0.3}
\end{figure}

Figure \ref{fig:samplesize-deviate_pH_0.3} displays the PCSs under both scenarios of interest ($\mathcal{S}_L$ and $\mathcal{S}_H$) when the trial is designed assuming $p_H=0.3$ and $\delta = 0.1$, across different target levels $\alpha_L$ and $\alpha_H$ and and four sample size deviation types. When the actual number of patients enrolled deviates from the planned sample size, the PCS remains close to the targeted levels.

\bibliographystyle{plainnat}


\end{document}